\documentclass[]{aa}

\usepackage{natbib}
\usepackage{breqn}
\usepackage{multirow}
\usepackage{hyperref} 
\usepackage{float}
\usepackage[dvipsnames]{xcolor}

\begin{document}

\title{MoBiDICT: new 3D static models of close, synchronized binaries in hydrostatic equilibrium}
\titlerunning{MoBiDICT}

\author{L. Fellay\inst{1} \and M.-A. Dupret\inst{1} }
\institute{STAR Institute, University of Liège, 19C Allée du 6 Août, B$-$4000 Liège, Belgium}
\date{December, 2022}

\abstract{In close binary systems, tidal interactions and rotational effects can strongly influence stellar evolution as a result of mass-transfer, common envelope phases, ... All these aspects can only be treated following improvements of theoretical models, taking into account the breaking of spherical symmetry occurring in close binaries.  Current models of binary stars are relying either on the so-called « Roche model » or the perturbative approach that in each case results on several assumptions concerning the gravitational,  tidal and centrifugal potentials.}
{Developing precise 3D models of stellar deformations and studying the robustness of the Roche and perturbative models in different deformation regimes.}
{We developed a new non-perturbative method to compute precise structural deformation of binary system in three dimensions that is valid even in the most distorted cases.  We then compared our new method to the Roche and perturbative models for different orbital separations and binary components.}
{We found that in the most distorted cases both Roche and perturbative models are significantly underestimating the deformation of binaries.  The effective gravity and the overall structural deformations are also noticeably different in the most distorted cases leading, for the interpretation of observations, to modifications of the usual gravity darkening generally obtained through the Roche model. Moreover we found that the dipolar term of the gravitational potential,  usually neglected by the perturbative theory,  has the same order of magnitude than the leading tidal term in the most distorted cases.}
{We developed a new method that is capable of precisely computing the deformations of binary system composed of any type of stars, even compact objects. For all  stars studied the differences in deformation with respect to the Roche or perturbative models  are significant in the most distorted cases impacting both the interpretation of observations and the theoretical structural depiction of these distorted bodies. In the weaker deformation regimes the Roche model is a viable option to study the surface properties of binaries while the perturbative model is strongly favoured to evaluate the structural deformations.}
\maketitle

\noindent

\section{Introduction}
The recent work of \cite{Sana2012, Sana2013,Sana2014} showed that a significant fraction of massive binary stars ($>70\%$) are impacted during their lifetime by interactions with their companions.  While massive stars have a high occurrence rate in binary system \citep{Sana2011,Sana2012, Sana2013,Sana2014}, both evolved or Main-Sequence (MS) low-intermediate mass stars are also commonly found in binary systems \citep{Slawson2011,Price-Whelan2018, GaiaBinaries2020}.
The recent Gaia Data Release 3 \citep{Gaia2020} offers the possibility to catalogue non-single stars as more than 800,000 of them were discovered \citep{GaiaBinaries2020} allowing to find and study the most peculiar systems.
\\~\\In a binary system, the presence of a nearby companion  is altering the classical evolutionary path of each binary component \citep{Hurley2002,Sana2012} leading in the most extreme cases to X-ray binaries \citep{Savonije1978,Patterson1984} or gamma Ray bursts \citep{Aharonian2005}, for example.  For any type of close binaries,  Roche Lobe (RL) overflow is a common phenomenon \citep{Sana2011,Sana2012, Sana2013,Sana2014} characterised by material and angular momentum transfer that is even stripping stars from their envelope for some systems and believed to be at the origin of hot subdwarfs \citep{Han2002, Han2003, Heber2009}.  Moreover,  phases of common envelope can allow the binary components to exchange angular momentum or chemicals, for example.  Before reaching such stages, stars are experiencing deformations induced by their mutual interactions dominated by the gravitational, tidal and centrifugal forces. 
\\~\\A correct modelling of these deformations is crucial to determine an accurate stellar structure of each component of the binary system during these phases and study their evolution.  These models are also necessary as stating points to study in more details the physics behind mass transfer that is currently relying on numerous free parameters and assumptions.  Such models can, finally, contribute to precisely compute system apsidal motions that impose direct observable constraints to the stellar structure in binary system usually theoretically obtained with the perturbative model \citep{Rosu2020}.
\\~\\The modelling of deformations is currently based on two independent types of models: the Roche model and the perturbative model.  
The principle of the Roche model is to treat binary components as point-mass bodies using a simplified expression of the gravitational and tidal potentials.  On the other hand,  the principle of perturbative models is to treat the centrifugal and tidal forces as a first order perturbation to the spherical symmetry, only accounting for the leading orders of gravitational potential \citep[$\ell=0,2$][]{Kopal1959, Kopal1978} and neglecting in particular the dipolar term ($\ell=1$).  Both models are first order approximations considering spherically symmetric stellar density distributions.    In the least distorted cases,  the different assumptions of classical modelling procedure are justified and allowing to study the stellar surface deformations, effective gravity or even structure \citep{Siess2013, Palate2013, PHOEBE2018, Fabry2022}. However, in the most distorted cases,  the redistribution of mass caused by the deformation as well as the different assumptions made by the classical models can become inaccurate to properly compute and study the stellar deformation,  in particular in the low-mass regime \citep{Landin2009}.
\\~\\To study the limits of classical models and get an accurate computation of the stellar surface deformations even in the most distorted cases,  we developed a new non-perturbative approach in three dimensions (3D) based on a method presented in \cite{Roxburgh2004, Roxburgh2006}  and recently used by \cite{Manchon2021,Rubis2023} to model rotating stars in two dimensions. This tool called Modelling  Binary Deformation Induced by Centrifugal and Tidal forces (MoBiDICT) is developed in Fortran 95 for the computational efficiency.  The principle of MoBiDICT is to iteratively solve Poisson's equation including the centrifugal and tidal forces in a non-perturbative way.  More than obtaining accurate surface quantities,  such non-nerturbative models are giving access to the entire, precise, 3D structure of each binary system component, allowing accurate computation of gravity darkening or apsidal motion constant even in the most distorted cases.  Finally,  the impact of deformations on the time evolution of the stellar internal structure will be accessible with such detailed models.
\\~\\In this article, we start by giving a technical and physical description of our new modelling method in Sect.~\ref{sec_stellar_models}. In Sect.~\ref{sec_post_treatment},  we present the post treatment of our code that computes the effective gravity and give the physical insights to couple MoBiDICT to classical 1D stellar evolution codes. Section~\ref{sec_comparison_Roche} is dedicated to the presentation of the Roche model and its comparison with our new models. The presentation and comparison with the perturabative model is done in Sect.~\ref{sec_comparison_perturbative}. Finally, the conclusion, the possible improvements and applications of our new types of models are discussed in Sect.~\ref{sec_conclusion}.

\section{Stellar models and properties}\label{sec_stellar_models}
In this section we describe and give the technical details of our novel modelling procedure to compute 3D non-perturbative models of close synchronised binaries in hydrostatic equilibrium. Our method is an iterative procedure designed to include the impact of stellar deformations on the stellar structure of the whole system. 
\\In Sect.~\ref{subsect_general_modelling},  we present the general properties of the problem that we solve,  as well as  the general principle of MoBiDICT. Then, in Sect.~\ref{subsec_psi_tot} and \ref{subsec_rho_tot}, we are going in the detail of the modelling procedure by describing the different steps of each iteration. Finally,  in Sect.~\ref{subsec_convergence_c}, we present our convergence criterion and the different readjustments made at the end of each modelling iteration.

\subsection{General modelling procedure}\label{subsect_general_modelling}
In the case of a binary system, the equations governing the hydrostatic structure of a given rotating component $i$
\begin{equation}\label{eq_hydrostatic_structure}
\frac{\boldsymbol{\nabla} P}{\rho}= -\boldsymbol{\nabla}\left(\Psi_1 +\Psi_2 + \Psi_{\rm{centri}}\right)=\textbf{g}_{\rm{eff}},
\end{equation}
and the Poisson equation:
\begin{equation}\label{eq_Poisson_classique}
\Delta \Psi_i = 4\pi G \rho_i,
\end{equation}
where $\rho(r,\mu,\phi), P(r,\mu,\phi)$ respectively denote the density and the pressure in the spherical polar coordinates $(r,\mu,\phi)$, with $\mu=\cos(\theta)$, $\theta$ being the polar angle, $\phi$ being the azimuthal angle and $r$ is the radial coordinate.  $\Psi_1(r,\mu,\phi)$ denotes the gravitational potential of the arbitrary selected primary star of the system, $\Psi_2(r,\mu,\phi)$ is the gravitational potential of its companion and $\Psi_{\rm{centri}}$ is the centrifugal potential later defined in Eq.~(\ref{eq_centri_pot}).  Equation~(\ref{eq_Poisson_classique}) is a second order non-linear differential equation in three dimensions with,  as boundary conditions, $\Psi_i(r,\mu,\phi)$ $\rightarrow 0$ at the infinity. 
\\~\\Assuming stars with a solid body rotation, all the forces here considered are derived from a potential thus the problem is conservative. By combining Eq.~(\ref{eq_hydrostatic_structure}) and its rotational:
\begin{multline}
\boldsymbol{\nabla}\times\left(\frac{\boldsymbol{\nabla} P}{\rho}\right)= -\frac{\boldsymbol{\nabla} \rho\times\boldsymbol{\nabla}P}{\rho^2}=-\boldsymbol{\nabla}\times\boldsymbol{\nabla}\left(\Psi_1 -\Psi_2 - \Psi_{\rm{centri}}\right)\\=\boldsymbol{0},
\end{multline}
we can deduce that the structure of each star is barotropic which means that the density and pressure are constant on a given equipotential. At a given chemical composition, the structure of stars becomes thermotropic implying that the temperature is also constant on the equipotentials.  To take advantage of these properties,  we assume that the density of each star is given by a one dimensional (1D) spherically symmetric input model along a given arbitrary direction ($\mu_{\rm{crit}}, \phi_{\rm{crit}}$).  Therefore,  in this direction we can associate a total potential to a density distribution and,  by computing the total potential on the whole star,  deduce the 3D density profile of each component of the system.  The choice of the direction ($\mu_{\rm{crit}}, \phi_{\rm{crit}}$) is detailed in Sect.~\ref{sec_mass_conservation}.
\\~\\By assuming that stars have the rotation axis aligned to the orbital rotational axis, two planes of symmetry are appearing: the first one is the orbital plane, the second one in the plane including the two stellar rotation axes and the orbital rotation axis. With these two symmetries, a description of one quarter of each star is sufficient to model the entire system in 3D.
\\~\\The principle of our modelling method is to iteratively solve Poisson's equation on a spherical harmonic basis for each star in order to compute, in a non-pertubative way, the impact of tidal and centrifugal forces on the stellar hydrostatic structure. This method is a generalisation of the method of \cite{Roxburgh2004, Roxburgh2006} in 3D and for binary systems. Each iteration of our modelling technique is divided as follows
\begin{enumerate}
\item as initial model we take an input density profile from a spherically symmetric 1D model: $\rho_i(r,\mu, \phi)=\rho_{i,  \mathrm{1D}}(r)$, obtained through a stellar evolution code for each star $i$ composing the system.
\item we solve Poisson's equation (Eq.~\ref{eq_Poisson_classique}) knowing $\rho_i(r,\mu, \phi)$ and we compute $\Psi_{\rm{tot}}= \Psi_1 + \Psi_2+ \Psi_{\rm{centri}}$ for each star.
\item we deduce $\rho_i(r,\mu, \phi)$ from $\Psi_{\rm{tot}}(r,\mu, \phi)$ using the barotropic property of our problem. More precisely,  along the direction ($\mu_{\rm{crit}}, \phi_{\rm{crit}}$) we compute $\Psi_{\rm{tot}}(r,\mu_{\rm{crit}}, \phi_{\rm{crit}})$,  and impose that $\rho_i(r,\mu_{\rm{crit}}, \phi_{\rm{crit}})=\rho_{i,  \mathrm{1D}}(r)$.  Since $\rho_i$ and $\Psi_{\rm{tot}}$ are two monotonic functions of $r$, we obtain the function $\rho_i$($\Psi_{\rm{tot}}$).  Finally, we simply have $\rho_i(r, \mu, \phi)$ = $\rho_i$($\Psi_{\rm{tot}}(r, \mu, \phi)$). 
\item we estimate the differences $\delta\rho_i(r,\mu, \phi)$ and $\delta\Psi_i(r,\mu, \phi)$ that we are using as convergence indicator and readjust the model. If the model has not converged we start back from the step $2$ using $\rho_i(r,\mu, \phi)$ obtained in step $3$ as an input.
\end{enumerate}

\subsection{Solution for $\Psi_{\rm{tot}}$ given $\rho_i$}\label{subsec_psi_tot}
Given $\rho_i(r,\mu, \phi)$ on a grid $(r,\mu, \phi)$ for each star $i$ we are going to solve the Poisson's equation (Eq.~\ref{eq_Poisson_classique}) in order to obtain the potential $\Psi_i(r,\mu, \phi)$ and deduce $\Psi_{\rm{tot}}= \Psi_1 + \Psi_2+ \Psi_{\rm{centri}}$.  We decided to adopt an individual spherical coordinates grid $(r_i,\mu, \phi)$ centred on each star $i$ with $r_i$ pointing toward the centre of mass of the system when $\phi=0$ and $\mu=0$. The colatitude $\mu$ and $\phi$ are sharing a common mesh for each star while $r_i$ is depending on the mesh of the input 1D models. Taking advantage of the two symmetrical plans of the problem, we can limit the angular domain to $\mu\in [1,0]$ and $\phi\in [0,\pi]$.
\\~\\Generalising the method presented by \cite{Roxburgh2006} in 3D and for binaries, $\rho$ and $\Psi$ are going to be expressed as a finite sum of spherical harmonics with 
\begin{equation}
\rho_i(r_i,\mu, \phi)=\sum^L_{\ell=0} \sum^{(\ell-p)/2}_{k=0}   \rho_{i, \ell}^m (r_i) P_{\ell}^m (\mu) \cos(m\phi),
\end{equation}
and,
\begin{equation}
\Psi_i(r_i,\mu, \phi)=\sum^L_{\ell=0} \sum^{(\ell-p)/2}_{k=0}   \Psi_{i, \ell}^m (r_i) P_{\ell}^m (\mu) \cos(m\phi),
\end{equation}
where $p=0$ if $\ell$ is even, $p=1$ otherwise, and $m=2k+p$. 
$L$ is the free parameter of our modelling defining the maximum degree of the spherical harmonics to account for.   $ P_{\ell}^m (\mu) \cos(m\phi)$ have been normalized such as
\begin{equation}
\int^{\pi}_0 \int^{1}_0 P_{\ell}^{m} (\mu) \cos(m\phi) P_{\ell^{\prime}}^{m^{\prime}} (\mu) \cos(m^{\prime}\phi) \mathrm{d}\mu \mathrm{d}\phi =\delta_{\ell \ell^{\prime}}\delta_{m m^{\prime}},
\end{equation}
where $\delta_{i j}$ is Kronecker's delta,  only $\neq 0$ and $=1$ when $i=j$.
$\Psi_{i, \ell}^m $ are going to be obtained by solving Poisson's equation on the spherical harmonics basis that is expressed as
\begin{equation}\label{eq_psi_lm}
\frac{1}{r_i^2}\frac{\mathrm{d}}{\mathrm{d}r_i}\left(r_i^2\frac{\mathrm{d} \Psi_{i, \ell}^m}{\mathrm{d}r_i}\right) - \frac{\ell(\ell+1)}{r_i^2} \Psi_{i, \ell}^m = 4 \pi G \rho_{i, \ell}^m,
\end{equation}
with the boundaries conditions: $\Psi_{i, \ell}^m(0)=0$ if $\ell \neq 0$ and 
\begin{equation}\label{eq_rho_boundaries}
(\ell+1)\Psi_{i, \ell}^m+r_i\frac{\mathrm{d}\Psi_{i, \ell}^m}{\mathrm{d} r_i} = 0$ at $r_i=R_{i,0},
\end{equation}
where $R_{i,0}$ denotes a given radius at the exterior of the star $i$. 
To project the density on the spherical harmonics basis we used the integral relation : 
\begin{equation}\label{eq_rho_lm}
\rho_{i, \ell}^m(r_i)=\int_0^1\int_0^\pi \rho_i(r_i,\mu, \phi) P_{\ell}^m (\mu) \cos(m\phi) \mathrm{d}\phi \mathrm{d}\mu.
\end{equation}
As mentioned by \cite{Roxburgh2006}, the computation of $\rho_{i, \ell}^m(r_i)$ through Eq.(\ref{eq_rho_lm}) using a detailed mesh can become a real problem  in terms of computation time and precision required. To reduce this problem, we decided to use a mesh of $\mu$ and $\phi$ following the Gauss-Legendre quadrature method. This method allows giving an exact solution to an integral of a polynomial of degree $2n-1$,  where $n$ is the number of points in the integration domain. Drastically reducing the number of points in grids while increasing the integral precision,  the Gauss-Legendre quadrature method is particularly efficient to compute integrals of functions as our spherical harmonics.
\\~\\An integral representation of the solution of Eq.~\ref{eq_psi_lm} with the boundary conditions presented in Eq.~\ref{eq_rho_lm} can be written as 
\begin{multline}\label{eq_poisson_solved}
\Psi_{i, \ell}^m (r_i)=r_{i}^\ell 4\pi G\int^{r_{i}}_{R_{0,i}} r_{i}^{\prime-(2\ell+2)}\left[\int^{r_{i}^{\prime}}_{0} r_{i}^{\prime\prime\ell+2} \rho_{i, \ell}^m(r_i^{\prime\prime}) \mathrm{d}r^{\prime\prime}\right] \mathrm{d}r^{\prime}
\\ -\frac{r_{i}^{\ell}}{R_{0,i}^{2\ell+1}}\frac{4\pi G}{2\ell+1}\int^{R_{0,i}}_{0} r_{i}^{\prime\ell+2} \rho_{i, \ell}^m(r_{i}^\prime) \mathrm{d}r^\prime.
\end{multline}
An equivalent representation of Eq.(\ref{eq_poisson_solved}) used by \cite{Kopal1959, Kopal1978,Fitzpatrick2012} for example is expressed as 
\begin{multline}\label{eq_poisson_solved_fidz}
\Psi_{i, \ell}^m (r_i)=-\frac{4\pi G}{(2\ell+1)r_{i}^{\ell +1}}\int^{r_{i}^{\prime}}_{0} r_{i}^{\prime\ell+2} \rho_{i, \ell}^m(r_i^{\prime})\mathrm{d}r^{\prime}
\\ -\frac{4\pi G r_{i}^{\ell}}{2\ell+1}\int^{R_{0,i}}_{r} r_{i}^{\prime 1-\ell} \rho_{i, \ell}^m(r_{i}^\prime) \mathrm{d}r^\prime.
\end{multline}
Both equivalent solutions were implemented in our code, the only difference between the two being related to the numerical treatment of the integrals in the limit $r_i\rightarrow0$.  To avoid further complications in this region,  we exploited the linearity of $\rho_{i, \ell}^m(r_i^2)/r_i^\ell$ close to the singularity with our trapezoidal integration method to increase the integration precision.  With both solutions the same results were obtained,  for all the work presented below we used the solution given by Eq.(\ref{eq_poisson_solved}) .
\\Inside MoBiDICT all the equations solved and quantities used are dimensionless to facilitate the computations. To nondimensionalize the system, we used constants of binary system, the different dimensionless variables of the system are expressed as
\begin{multline}
\;\; \;\; \;\;\;\;\;\;\; x_i=r_i/a;\;\;\;\; m_i^'=m_i/ M_{\rm{tot}};\;\;\;\;   \rho_i^'=\rho_i/\left( \frac{M_{\rm{tot}}}{4\pi a^3} \right); \\ \Psi_i^'=\Psi_i/\left( \frac{GM_{\rm{tot}}}{a} \right);\;\;\;\;\Omega_{\star}^{'2}=\Omega_{\star}^2/\left( \frac{GM_{\rm{tot}}}{a^3} \right),
\end{multline}
where $a$ is the separation of the binaries, $M_{\rm{tot}}$ the total mass of the binary system,  $\Omega_{\star}$ the orbital rotation rate and $G$ the gravitational constant.  From a numerical point of view,  having dimensionless variables is useful. In our case, the quantities used to nondimensionalize were chosen to simplify the $4\pi G$ constant in Poisson's equation (Eq.~\ref{eq_Poisson_classique}) and the integral representation of its solution (Eqs.~\ref{eq_poisson_solved} and  \ref{eq_poisson_solved_fidz}). 
In the rest of this article we will keep the equations and quantities presented with dimensions to keep the physical sense of the problem.
\\~\\Using all the previously presented method, the algorithm adopted to compute the total potential of each star follows
\begin{enumerate}
\item with the density provided of each star $\rho_i(r_i,\mu, \phi)$ we determine their spectral expansion $\rho_{i, \ell}^m(r_{i})$ through Eq.~\ref{eq_rho_lm}. 
\item we numerically compute Poisson's equation (Eq.~\ref{eq_psi_lm}) solution given by Eq.~\ref{eq_poisson_solved} for each star to obtain their gravitational potential projected on the spherical harmonics basis:  $\Psi_{1, \ell}^m(r_1)$ and $\Psi_{2, \ell}^m(r_2)$.
\item using the proper coordinates conversion and the previously determined spectral quantities, namely $\Psi_{1, \ell}^m(r_1)$ and $\Psi_{2, \ell}^m(r_2)$, we determine each gravitational potential on the grid of each star obtaining $\Psi_1(r_1,\mu, \phi)$, $\Psi_1(r_2,\mu, \phi)$, $\Psi_2(r_1,\mu, \phi)$, $\Psi_2(r_2,\mu, \phi)$. Finally, we compute the total potential 
$\Psi_{\rm{tot}}= \Psi_1 + \Psi_2+ \Psi_{\rm{centri}}$ on the grid of each star.
\end{enumerate}
The centrifugal potential of each star, assuming a synchronized solid body rotation aligned to the orbital rotation axis, can be expressed as
\begin{multline}\label{eq_centri_pot}
\Psi_{\rm{centri}}(r,\mu,\phi)=-\frac{\Omega_{\star}^2}{2}\varpi^2(r,\mu,\phi)\\=-\frac{\Omega_{\star}^2}{2} \left[ \left( x_{\rm{CM}}-r\sin(\theta)\cos(\phi)\right)^2 + \left( r\sin(\theta)\sin(\phi)\right)^2\right],
\end{multline}
where $\varpi(r,\mu,\phi)$ is the distance, on the orbital plane, from the orbital rotation axis of the element considered, $\Omega_{\star}$ is the stellar and orbital rotation rate, $x_{\rm{CM}}$ is the distance of the star  from the centre of mass of the system.
\\Our formalism is not limited to solid body rotation with aligned rotational axis, by modifying the centrifugal potential we can extend the method to model the cases of non-synchronized, non-aligned solid body rotations.  These specific cases could be important to model,  since we expect non-synchronized stellar rotations close to the critical velocity as a result from mass transfer within binary systems. This generalisation of our formalism will be presented in a forthcoming article. 

\subsection{Solution for $\rho$ given $\Psi_{\rm{tot}}$}\label{subsec_rho_tot}
Given $\Psi_{\rm{tot}}$ that we determined in the Sect.~\ref{subsec_psi_tot} we want to estimate the corresponding density for of each star. We exploit the fact that the structure of each star is barotropic,  meaning that the density is constant on the equipotentials. 
\\Assuming that the density of a star $i$ along a given direction $\mu_{\rm{crit}}$ and $\phi_{\rm{crit}}$ is the density of the corresponding averaged 1D input model,  we can determine the functions $\rho_i(\Psi_{i, \rm{tot}}(r_i, \mu_{\rm{crit}}, \phi_{\rm{crit}}))=\rho_{i, \rm{1D}}(r_i)$ for each star. Then by simple function composition and interpolation we can estimate the density of the entire star with the total potential on each point of the grid.

\subsection{Readjustment and criterion for the convergence of the method}\label{subsec_convergence_c}
A readjustment of the density is necessary to conserve the total mass of each star and the system. By the simple integration of $\rho_{i, \ell=0}^{m=0}(r_i)$ we obtain the total mass of each star with our new density profiles, we then correct the entire stellar density profiles by the ratio of the mass from the average 1D input models to the new mass obtained.
\\~\\At this modelling step several readjustments are made in order to verify different basic laws impacted by the non-sphericity of the models.  In particular a system of equations verifying the balance of forces at the centre of each star is solved in order to properly recompute the centre of mass and the orbital period of the system to maintain the balance of the forces.  The system of equations solved at this modelling step is presented in Appendix~\ref{annexe_period_readjustment}.
\\~\\The convergence of our method is guided by the variations of the density and gravitational potential between two successive iterations. These variations are quantifying the contribution of each iteration to the deformation of each point of a star thus giving a direct control of the convergence of our modelling. The point at which deformation is expected to be the most important is the closest one to the Lagrangian point L1 in each star $i$: $(R_{s,i},\mu=0,\phi=0)$,  thus,  if the variation of potential and density at this point is negligible compared to the previous iteration,  then the entire model has converged. In the rest of our work $R_{s,i}$ denotes the radial coordinate of the stellar surface in a given direction($\mu,\phi$).  Our normalized convergence criterion for a model at the iteration $j$ is
\begin{equation}
\delta\rho_{i, j, \ell}^m=\frac{\rho_{i, j}(R_{s,i,j},0,0)- \rho_{i, j-1}(R_{s,i,j-1},0,0) }{\rho_{i, j}(R_{s,i,j},0,0)},
\end{equation}
or,
\begin{equation}
\delta\Psi_{i, j, \ell}^m=\frac{\Psi_{i, j}(R_{s,i,j},0,0)- \Psi_{i, j-1}(R_{s,i,j-1},0,0) }{\Psi_{i, j}(R_{s,i,j},0,0)}.
\end{equation}
Given the normalized precision that we have with our 1D average input models of about $10^{-8}$ on the density we used as criterion for convergence $\delta\rho_{i, j, \ell}^m<10^{-8}$ and $\delta\Psi_{i, j, \ell}^m<10^{-8}$ for each star. This value can be adjusted in order to reduce the computation time of our models if a high precision modelling is not required.
\section{Post-treatment of the models}\label{sec_post_treatment}
Once we obtained a final model with the desired precision,  the models are sent to a post processing,  phase during which the different properties of the models are extracted and used to potentially include our new 3D models in stellar evolutionary models using the method of \cite{Kippenhahn1970}.  Later revisited by \cite{Meynet1997},  this method is extensively used in the literature \citep{Roxburgh2004,Landin2009, Siess2013, Palate2013,  Fabry2022} to couple deformations to 1D stellar evolutionary models.
\\Following this method,  the stellar structural equations are averaged over isobars of our 3D models to give a one dimensional description of the stellar distorted structure usable by stellar evolution codes. 
\subsection{Mass conservation equation}\label{sec_mass_conservation}
Let us introduce $m_{\rm{p}}$ denoting the mass encompassed under an isobar of pressure $p$ and $r_{\rm{p}}$ the averaged radius such as $V_{\rm{p}}=\frac{4}{3}\pi r_{\rm{p}}^3$ is the volume under this isobar. 
\\With this notation, the mass conservation equation is unmodified and given by
\begin{equation}\label{eq_mass_conservation}
\frac{\mathrm{d} m_{\rm{p}}}{\mathrm{d} r_{\rm{p}}}= 4\pi\rho r_{\rm{p}}^2.
\end{equation}
By construction, the critical direction  ($\mu_{\rm{crit}}, \phi_{\rm{crit}}$) is chosen to verify Eq.~\ref{eq_mass_conservation} at each layer of the star implying that $ r_{\rm{p}, i}=r_i$ and $m_{\rm{p},i}=m_i$ in the whole star.

\subsection{Hydrostatic equilibrium}
Following Eq.~\ref{eq_hydrostatic_structure}, the infinitesimal distance $\mathrm{d}n$ between the equipotentials $\Psi_{\rm{tot}}$ and $\Psi_{\rm{tot}}+ \mathrm{d} \Psi_{\rm{tot}}$ is given by 
\begin{equation}\label{eq_dn}
\mathrm{d}n=\vert\textbf{g}_{\rm{eff}}\vert^{-1} \mathrm{d} \Psi_{\rm{tot}}.
\end{equation}
With this formalism, the volume between two isobars is obtained with
\begin{equation}\label{eq_volume_between}
\mathrm{d} V_{\rm{p}} = \int_{\Psi_{\rm{tot}}} \mathrm{d}n \mathrm{d}\sigma,
\end{equation}
where the integrals on ${\Psi_{\rm{tot}}}$ are denoting integrals taken over isobars of constant ${\Psi_{\rm{tot}}}$,  $\mathrm{d}\sigma$ is an isobar surface element expressed as
\begin{equation}\label{eq_sigma_expression}
\mathrm{d}\sigma=-r(\mu,\phi)^2  \frac{\vert\textbf{g}_{\rm{eff}}\vert \cdot \vert\textbf{r}\vert}{\textbf{g}_{\rm{eff}}\cdot\textbf{r}} \mathrm{d}\mu \mathrm{d}\phi.
\end{equation}
In Eq.~\ref{eq_sigma_expression}, $\vert g_{\rm{eff}}\vert$ denotes the norm of the effective gravity presented in Eq.(\ref{eq_hydrostatic_structure}) and $r(\mu,\phi)$ is the radial position of the chosen isobar in the direction $(\mu,\phi)$. The analytical expression of the effective gravity in our new models is developed in  Appendix \ref{annexe_effective_gravity}.
Equation~\ref{eq_volume_between} can be integrated and combined to Eqs.~\ref{eq_dn} and \ref{eq_sigma_expression} to obtain the volume under an isobar:
\begin{align}
V_{\rm{p}}  &= \nonumber  \int^{\Psi_{\mathrm{tot}}}_{\Psi_{\mathrm{tot, c}}}d \Psi_{\rm{tot}} \int_{\Psi_{\rm{tot}}}  \vert\textbf{g}_{\rm{eff}}\vert^{-1} \mathrm{d}\sigma\\ \ &=  \int_{\mu, \phi} \mathrm{d}\mu \mathrm{d}\phi \int^{\Psi_{\mathrm{tot}}}_{\Psi_{\mathrm{tot, c}}} r(\mu,\phi)^2 \dfrac{\partial r}{\partial \Psi_{\mathrm{tot}}} \mathrm{d}\Psi_{\mathrm{tot}} \\ \ &=\frac{1}{3}  \int_{\mu, \phi}  r(\mu,\phi)^3  \mathrm{d}\mu \mathrm{d}\phi\nonumber,
\end{align}
where $\Psi_{\mathrm{tot, c}}$ is the total potential at the centre of the star. Finally, the mass between two isobars is expressed as
\begin{equation}
\mathrm{d} m_{\rm{p}}=\rho\int_{\Psi_{\rm{tot}}} \mathrm{d}n \mathrm{d}\sigma =\rho \mathrm{d} \Psi_{\rm{tot}} \int_{\Psi_{\rm{tot}}} \vert\textbf{g}_{\rm{eff}}\vert^{-1} \mathrm{d}\sigma.
\end{equation}
Therefore,   the equation of hydrostatic equilibrium can be expressed as
\begin{equation}
\frac{\mathrm{d}P}{\mathrm{d}m_{\rm{p}}}=\frac{\mathrm{d}P}{\mathrm{d} \Psi_{\rm{tot}}}\frac{\mathrm{d} \Psi_{\rm{tot}}}{\mathrm{d}m_{\rm{p}}}=-\left( \int_{\Psi_{\rm{tot}}} \vert\textbf{g}_{\rm{eff}}\vert^{-1} \mathrm{d}\sigma \right)^{-1},
\end{equation}
that can be written as
\begin{equation}
\frac{\mathrm{d}P}{\mathrm{d}m_{\rm{p}}}=-f_{\rm{p}} \frac{G m_{\rm{p}}}{4\pi r_{\rm{p}}^4},
\end{equation}
where,
\begin{equation}\label{eq_fp}
f_{\rm{p}}=\frac{4\pi r_{\rm{p}}^4}{G m_{\rm{p}}}\left(\int_{\Psi_{\rm{tot}}} \vert\textbf{g}_{\rm{eff}}\vert^{-1} d\mathrm{d}\sigma \right)^{-1}.
\end{equation}
\subsection{Transport of energy}
In this part, we assume that the chemical composition of each star is constant on each isobar,  therefore the isobars are becoming isotherms.
\\In the radiative zone, the norm of the local flux is given by 
\begin{equation}\label{eq_flux_transfer_equation}
F=-\frac{4 a c T^3}{3 \kappa \rho}\frac{\mathrm{d}T}{\mathrm{d}n},
\end{equation}
where $T$ is the temperature of a given isobar,  $\kappa$ its opacity,  $a$ the radiation constant and $c$ the speed of light in the vacuum. Equation (\ref{eq_flux_transfer_equation}) can be rewritten using Eq.~\ref{eq_dn} and the mass conservation equation (Eq.~\ref{eq_mass_conservation})  as 
\begin{equation}\label{eq_gravity_darkening}
F=-\frac{4 a c T^3}{3 \kappa}\frac{\mathrm{d}T}{\mathrm{d}m_{\rm{p}}} \left( \int_{\Psi_{\rm{tot}}} \vert\textbf{g}_{\rm{eff}}\vert^{-1} \mathrm{d}\sigma \right)\vert\textbf{g}_{\rm{eff}}\vert,
\end{equation}
that is the classical formulation of the gravity darkening. 
\\By integrating on an entire equipotential we obtain the luminosity:
\begin{equation}
L_{\rm{p}}= -\frac{4 a c T^3}{3 \kappa}\frac{\mathrm{d}T}{\mathrm{d}m_{\rm{p}}}  \int_{\Psi_{\rm{tot}}} \vert\textbf{g}_{\rm{eff}}\vert^{-1} \mathrm{d}\sigma  \int_{\Psi_{\rm{tot}}} \vert\textbf{g}_{\rm{eff}}\vert \mathrm{d}\sigma.
\end{equation}
The temperature gradient can thus be expressed as 
\begin{align}
\frac{\mathrm{d}T}{\mathrm{d}m_{\rm{p}}} &= - \frac{3 \kappa L_{\rm{p}}}{4 a c T^3} \left( \int_{\Psi_{\rm{tot}}} \vert\textbf{g}_{\rm{eff}}\vert^{-1} \mathrm{d}\sigma  \int_{\Psi_{\rm{tot}}} \vert\textbf{g}_{\rm{eff}}\vert \mathrm{d}\sigma\right)^{-1}\\ \ &=- f_{\rm{T}} \frac{3 \kappa L_{\rm{p}}}{64\pi^2 r_{\rm{p}}^4 a c T^3},
\end{align}
where we introduced the corrective factor:
\begin{equation}\label{eq_ft}
f_{\rm{T}}= 16\pi^2 r_{\rm{p}}^4 \left( \int_{\Psi_{\rm{tot}}} \vert\textbf{g}_{\rm{eff}}\vert^{-1} \mathrm{d}\sigma  \int_{\Psi_{\rm{tot}}} \vert\textbf{g}_{\rm{eff}}\vert d\sigma\right)^{-1}.
\end{equation}
Finally the correction of the radiative gradient is 
\begin{equation}
\nabla_{\rm{rad, p}}=\frac{f_{\rm{T}}}{f_{\rm{p}}} \nabla_{\rm{rad}}
\end{equation}
In order to find the boundaries of convective zones, this corrective factor has to be applied to the radiative gradient. We can note that the 3D deformations computed are mainly significant in the upper regions of stars which means that $f_{\rm{T}}/{f_{\rm{p}}}\approx1$ close to the convective boundaries for most of the stars.

\subsection{Conservation of energy}
The energetic balance of a shell of mass $m_{\rm{p}}$ is trivial and give the classical equation:
\begin{equation}
\frac{\mathrm{d} L_{\rm{p}}}{\mathrm{d} m_{\rm{p}}}= \epsilon - T\frac{\mathrm{d}s}{\mathrm{d}t},
\end{equation}
where $\epsilon$ denotes the net energy production rate and $s$ is the entropy.
\\We can note that to couple MoBiDICT to 1D stellar evolution codes the computation of the effective gravity is required, the analytical expression of this effective gravity in our new models is given in the Appendix \ref{annexe_effective_gravity}.  Other main requirements are the detection of the equipotential and the interpolation of different quantities on each equipotential,  assuming that the potential of each star in the direction ($\mu_{\rm{crit}}, \phi_{\rm{crit}}$) is corresponding to the reference 1D stellar input model.
\\In this work $f_{\rm{T}}$ and $f_{\rm{p}}$ are also employed as structural deformation indicators to study the differences between  models as these quantities are isobar averages of the stellar deformation over the entire stellar structure.

\section{Comparison with the Roche model }\label{sec_comparison_Roche}

One classical way of describing the total potential in a binary system is the Roche model,  while this model is making strong assumptions on the stellar structure,  it also allows to have simple analytical expressions of the total potential at a given point of the system.
\\In this Section,  we are explore the Roche model looking in particular for the differences between the Roche model and our more accurate models.  Section~\ref{subsec_formulation_roche} is dedicated to a theoretical presentation of the Roche model and the different assumptions made by this model.  In Section~\ref{subsec_type_of_stars_studied},  we present the different stellar models and binary systems that will be modelled and compared in the following sections. In Sect.~\ref{subsec_surface_deformation_roche},  we study the difference in surface deformation between the Roche model and our models for different types of stars.  Section~\ref{subsec_surface_gravity_roche} is focussed on the differences of surface gravity that are essential for the computation of the gravity darkening and therefore to interpret observations of binary systems.  Finally, in Sect.~\ref{subsec_structural_deformation_roche},  we explore the differences of structural modelling between the Roche model and our modelling for different types of stars composing the binaries. 
\subsection{Theoretical formulation of the Roche model }\label{subsec_formulation_roche}
The Roche potential describes the total potential on a given point of the system that is produced by two point-like gravitationally interacting bodies and the consequent centrifugal potential created.  The main advantage of the Roche model is the possibility to have simple analytical expression of the total potential. To obtain such expressions several assumptions are made: the orbit of the stars are assumed circular and the stellar rotation synchronized to the orbital rotation,  the stars are assumed to behave as point like,  which means that the mass of their  envelope is neglected to compute the tidal potential,  the stars are also assumed to have a spherically symmetric density profile even in the most distorted cases.
\\The Roche model is overly used in the literature to compute the gravitational potential of distorted binaries at their surface and outside of stars \citep{Siess2013, Palate2013,  Fabry2022}.  In the most extreme cases this potential is even used to compute structural deformation of the components of binary systems through the coefficients $f_{\rm{p}}$ and $f_{\rm{T}}$ presented below.
\\In its classical formulation the total Roche potential of a given point in the grid of the arbitrary chosen primary of the system is given by 
\begin{equation}\label{eq_pot_roche_tot}
\Psi_{\rm{tot, i}}=\Psi_{1} + \Psi_{2} + \Psi_{\rm{centri}},
\end{equation}
where $\Psi_{1}$ and $\Psi_{2} $ are,  respectively,  the gravitational potential of the primary and secondary,  expressed as
\begin{equation}\label{eq_pot_roche_simple}
\Psi_{i} = -\frac{Gm_i(r_i)}{r_i}\ \rm{ or }\ \Psi_{i} = -\frac{GM_i}{r_i},
\end{equation}
depending on the author and the utilisation of the models.
The same assumptions than in our code are made concerning the circular orbits and the synchronized rotation, thus $\Psi_{\rm{centri}}$ is given by Eq.\ref{eq_centri_pot}.
\\The gravitational potential inside a star given by Roche model as expressed in Eq.~\ref{eq_pot_roche_simple} is not a solution of Poisson's equation (this can be directly seen by inserting Eq.~\ref{eq_pot_roche_simple} in Eq.~\ref{eq_Poisson_classique}).  Therefore,  using this formulation of the Roche potential to correct a stellar structure  at the hydrostatic equilibrium as done in Sect.~\ref{sec_post_treatment} can be problematic.  To solve this particular issue we derived a "refined" version of the Roche model,  guaranteeing that the gravitational potential is a solution of Poisson's equation while keeping it untouched outside of the star and expressed as in Eq.~\ref{eq_pot_roche_simple}. This version of the Roche potential that we call "refined Roche" in the rest of this article is expressed as
\begin{equation}
\Psi_{i} =\int_{R_{0, i}}^{r_i }\frac{G m_i(r)}{r^2} \mathrm{d}r -\frac{GM_i}{R_{0, i}}.
\end{equation}
In the following we are going to focus on the differences in surface and structural deformations as well as surface gravity obtained by the classical Roche model, by our refined Roche model and our non-perturbative modelling done with MoBiDICT.

\subsection{Types of stars studied}\label{subsec_type_of_stars_studied}

\begin{figure}[h]
\centering
\includegraphics[width=\hsize]{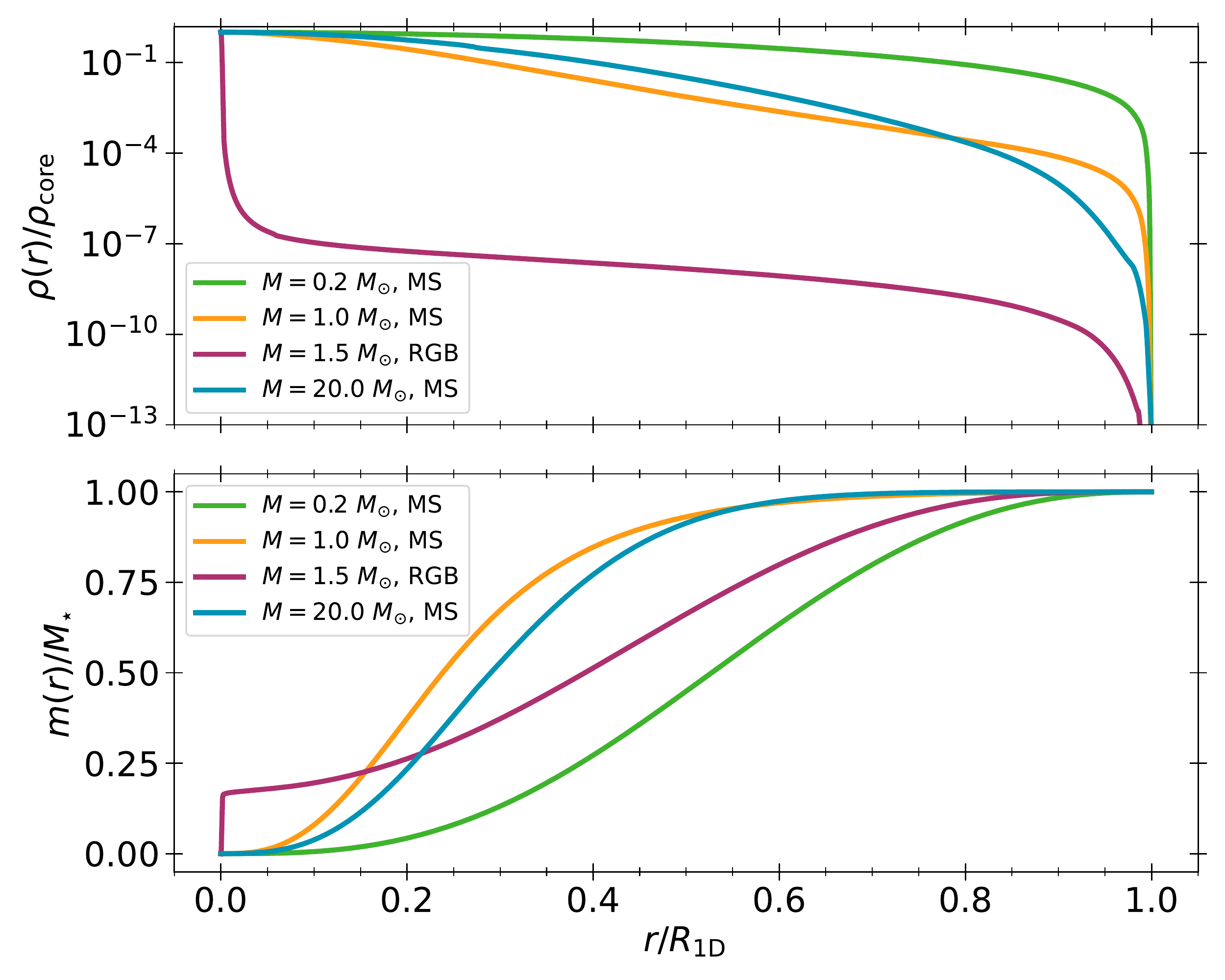}
\caption{1D input density profile normalized by the core density and mass profile of each component of the twin binary systems studied in this article. The upper panel corresponds to the stellar density profile while the bottom panel corresponds the mass profile normalized by the total mass of each star.  }
\label{fig.1D_density_profile}
\end{figure}

The surface deformation given by our models and the Roche model are studied in this work for several types of stars.  We made four 1D stellar models with the Code Liegeois d'Evolution Stellaire \citep[CLES,][]{Scuflaire2008a}  to represent a diversity of types of stars studied and compared  in the work. The stellar properties of each model is presented in Tab.~\ref{table_stella_models}.  Our method is not limited to these particular types of stars or binaries, as long as a 1D density profile of each star composing the system can be provided, its deformation can be computed.  In most extreme cases, our method is even able to model the deformations of systems with compact objects such as white dwarfs.
\begin{table}[h]
\centering
\caption{Summary of the stellar properties of the 1D input models used in this work.}\label{table_stella_models}
\resizebox{\hsize}{!}{\begin{tabular}{llllllll}
\hline\hline
\multirow{2}{*}{Stellar parameters} & \multirow{2}{*}{Model 1} &  \multirow{2}{*}{Model 2} &  \multirow{2}{*}{Model 3} &  \multirow{2}{*}{Model 4} \\
                                    &                                                          &                                                          &                                                         &                       \\ 
\hline
Mass {[}$M_\odot${]}                & $0.2$                              & $1.0$   & $1.5$ & $20.0$          \\
Radius {[}$R_\odot${]}              & $0.22$                               & $1.03   $  & $11.1$ & $6.01 $          \\
Age {[}Gyr{]}                     & $2.0$                               & $2.0 $& $2.0  $ & $0.002  $      \\
Evolutionary stage        & MS & MS & RGB                         & MS         \\
Effective temperature {[}$K${]}     & $3342 $                                   & $6080$    & $4621  $    & $35\ 859 $            \\
Luminosity {[}$L_\odot${]}          & $0.005$                              & $1.31$  & $51.1$ & $ 53\ 921 $         \\
Metallicity $Z_0/X_0$           & $0.014 $                             & $0.014 $  & $0.014 $ & $0.014 $         \\
Core Hydrogen $X_{\rm{c}}$           & $0.693$                             & $0.469 $  & $0 $ & $0.582 $         \\
\hline
\end{tabular}
}
\end{table}
\\~\\In this article we focused on twin systems,  with two identical stars, to have a point of reference to compare the distortions of different stellar models as the deformation of a binary system component is depending on the stellar structure of its companion.  However,  MoBiDICT works also perfectly with 2 non-identical components.  In MoBiDICT,  the type of star studied is only defined by the input density profile of each star then associated to a total potential along the direction ($\mu_{\rm{crit}}$,  $\phi_{\rm{crit}}$).  We illustrated the different 1D density profiles normalized by the core density and the mass profiles in function of the normalized radius of each star studied in this article in Fig.~\ref{fig.1D_density_profile}. 
\\~\\Two groups of models can be distinguished in Fig.~\ref{fig.1D_density_profile}, models with high or low envelope mass fractions. While massive and solar-like stars have an envelope mass nearly negligible, low-mass and RGB stars have most of their masses located in their envelope. In particular about $80\%$ of our RGB star mass is located in its convective envelope,  the transition from the radiative core to the convective envelope being located at $r/R_{\mathrm{1D}}\sim 0.05$ for this star.
\\In the following sections we compare the deformations obtained with the different types of stars presented above with the Roche models and with our new type of models. We then relate those differences directly to their 1D density and mass profiles presented in Fig~\ref{fig.1D_density_profile}..

\subsection{Surface deformation}\label{subsec_surface_deformation_roche}
\begin{figure}[h]
\centering
\includegraphics[width=\hsize]{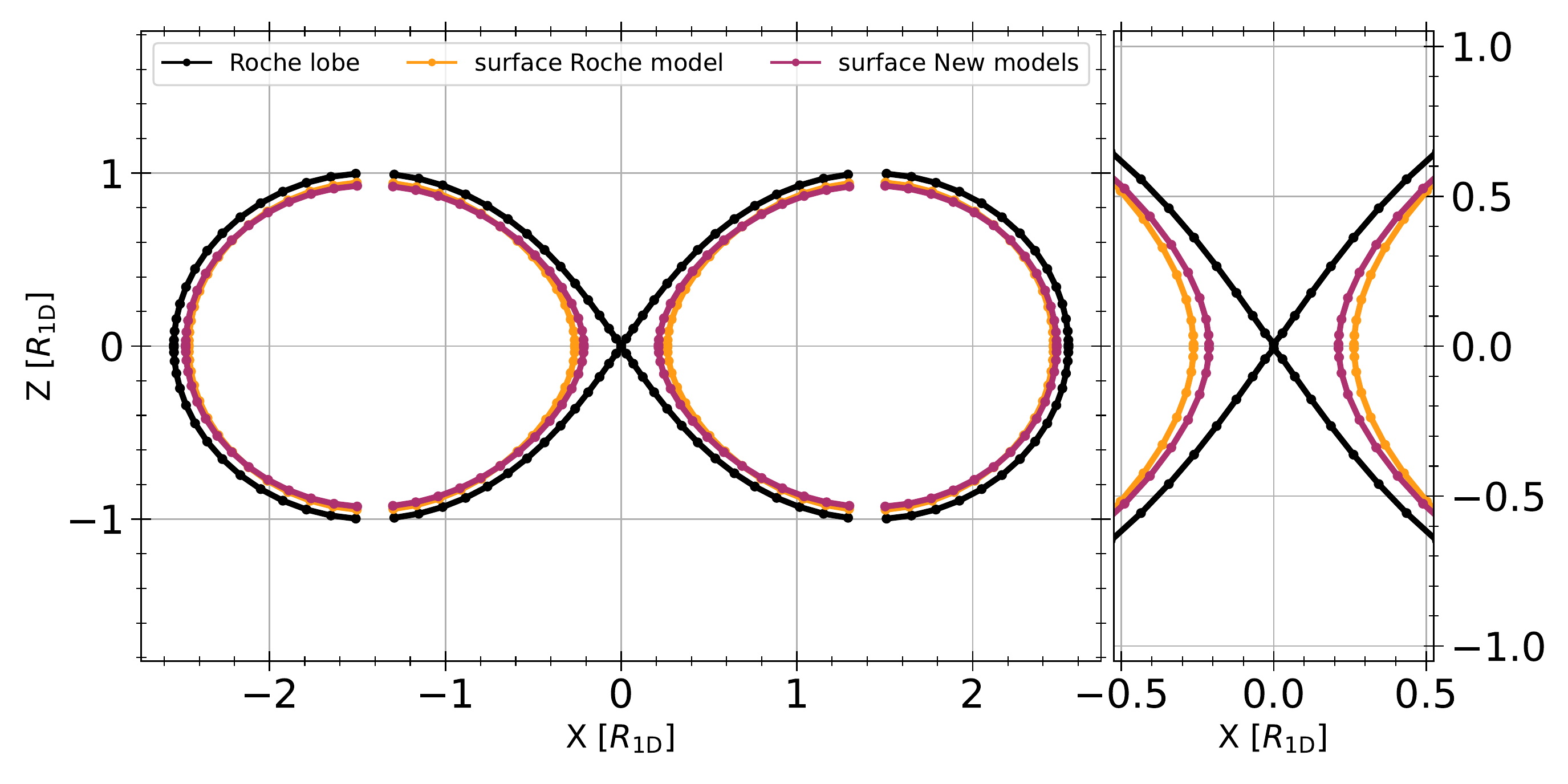}
\caption{ Surface deformation of the twin binary system of our $0.2$ $M_\odot$ stars with a separation of $2.8\ R_{\rm{1D}}$  (corresponding to a period of 2 hours and 11 minutes) as viewed from the side of the system. The black curve corresponds to the Roche lobe of the system, the yellow curve is the surface of each star given by the Roche model while the purple curve corresponds to the surface of each star given by the modelling with MoBiDICT.  In this particular case, the filling of the Roche lobes is of $81.7\%$ with each model. The absence of points near $\mu=1$ is explained by the Gauss-Legendre quadrature method that is creating no grid points in this region.}
\label{fig.surface1M0}
\end{figure}

\begin{figure}[h]
\centering
\includegraphics[width=\hsize]{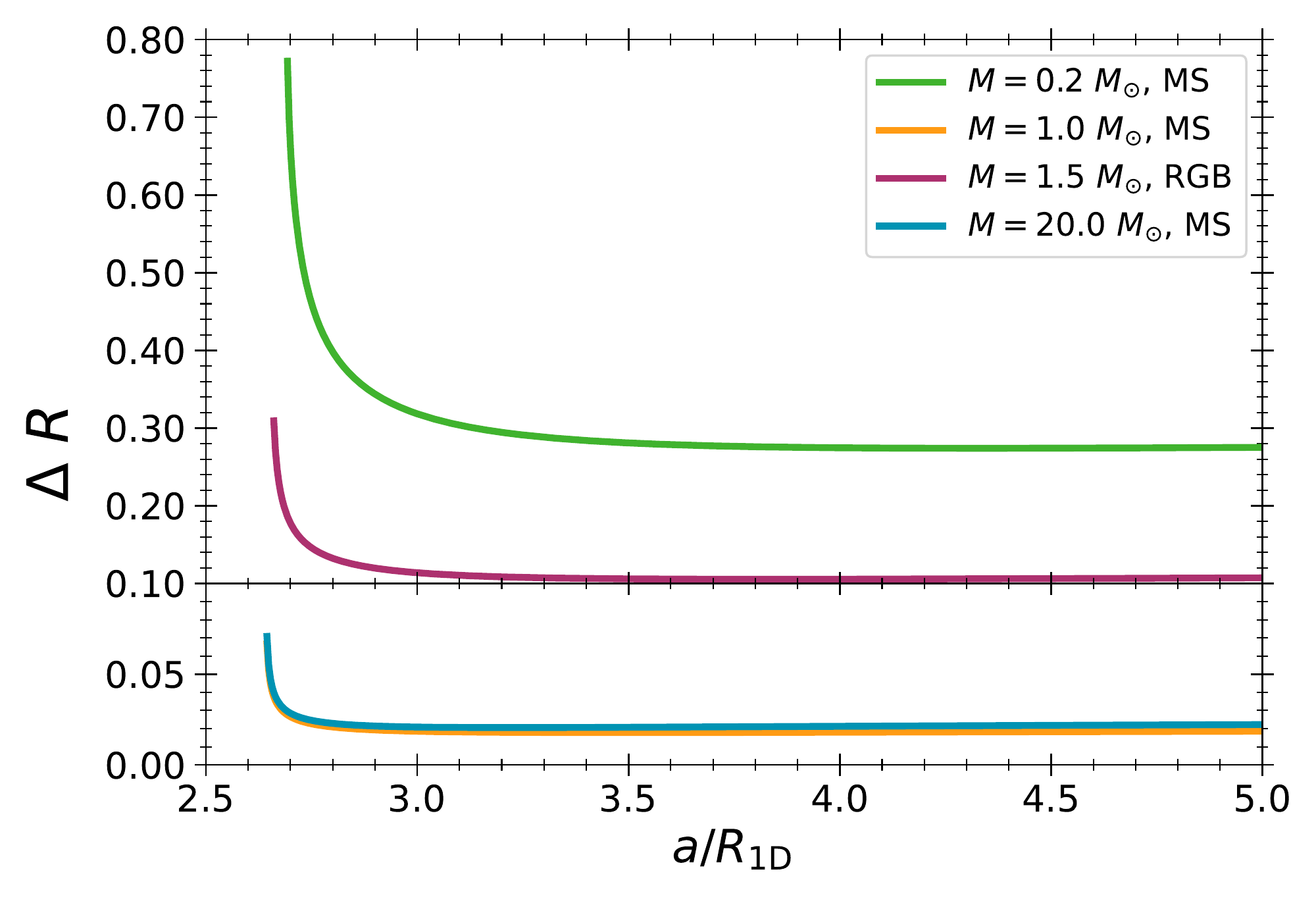}
\caption{ Difference in deformation at the closest point $\Delta R$ between Roche and our models as a function of the orbital separation of the components normalized by the input 1D stellar radius $a/R_{\rm{1D}}$. }
\label{fig.surface_deff_R_all_models}
\end{figure}

The surface deformation of binary stars is directly related to the total potential of the system.  In our models the surface of a star composing the binary system is defined by the equipotential corresponding to the photosphere of our 1D input stellar model in the direction ($\mu_{\rm{crit}},\phi_{\rm{crit}}$) as presented in Sect.~\ref{subsec_rho_tot}. 
\\Both classical Roche models and our refined version will give the same deformations at the surface as the total potentials have the same expressions outside of the star.  In order to study the deformation given by the Roche model we are going to assume, as done in our models, that the volumetric average radius is coinciding with the 1D input average radii meaning that the volume and mass of the models are conserved. Using this property,  we can define the surface given by the Roche model by the same criterion than with our models.
\\The position of the surface as given by the Roche models and our model are compared in the case of the $0.2$ $M_\odot$ with an orbital separation of $2.8\ R_{\rm{1D}}$ star in Fig.~\ref{fig.surface1M0}.  $R_{\rm{1D}}$ denotes the radius of the 1D spherically symmetric model.
\\Figure~\ref{fig.surface1M0} shows a visible difference in surface deformation between our models and the Roche model. To quantify this difference we introduce a new quantity:
\begin{equation}
\Delta R=\frac{R_{\rm{s, MoBiDICT}}(\mu=0, \phi=0)-R_{\rm{s, Roche}}(\mu=0, \phi=0)}{R_{\rm{s, Roche}}(\mu=0, \phi=0)-R_{\rm{1D}}},
\end{equation}
that is the difference in deformation at the closest point normalized by the deformation between the Roche model and the 1D input model. For the system presented in Fig.~\ref{fig.surface1M0},  this factor $\Delta R=0.368$,  which means that with MoBiDICT we have 36.8$\%$ more deformation than the Roche model at the closest surface point to the Lagrangian point L1. The difference of surface deformation is a function of the orbital separation between the two components of the binary system. To have a comparison  between different types of stars,  we are using the ratio of orbital separation to initial 1D stellar radius $a/R_{\rm{1D}}$ as indicator of the orbital separation with respect to the type of stars in the system. 
\\While in the rest of this deformation study we focussed on the most distorted region,  Fig.~\ref{fig.surface1M0} also shows that surface deformation discrepancies are appearing at the poles, where our model is more contracted than Roche's.
\\To study the significance of the modifications to the surface deformation obtained with MoBiDICT for the different types of stars,  we computed a grid of deformed models of twin binary systems composed of each stellar type presented previously. For each system,  we varied the orbital separation to explore the dependency of the deformation discrepancy to the orbital separation of the components. The results obtained with this grid of models are presented in Fig.~\ref{fig.surface_deff_R_all_models}.
\\~\\Figure~\ref{fig.surface_deff_R_all_models} shows that the differences of deformations are extremely dependent on the type of star studied. 
\\In the case of a massive and solar like MS stars, Fig.~\ref{fig.surface_deff_R_all_models} shows minor discrepancies of surface deformation even when the stars are close to fill their Roche lobes. At most, the deformation discrepancy reaches $7\%$ before a common envelope or mass transfer phases.
\\As showed in Fig.~\ref{fig.surface_deff_R_all_models},  RGB stars are significantly more deformed with our modelling (up to $\sim35\%$) than with the Roche model when $a/R_{\rm{1D}}\lesssim 3$.   Moreover, this discrepancy remains significant even at large $a/R_{\rm{1D}}$ with an asymptotic value of $\Delta R=0.10$. 
\\Finally,  it is for low mass stars that the corrections are the most significant.  We can see in Fig.~\ref{fig.surface_deff_R_all_models} that, when $a/R_{\rm{1D}}\lesssim 3.5$,  the Roche model totally underestimates the deformations up to $\sim 80\%$ just before reaching a common envelope or mass transfer phase. 
\\~\\One would expect for RGB stars that our new modelling does not significantly change the surface deformation obtained with the Roche model  as RGB are evolved stars with a high density contrast between the radiative core and convective envelope,  justifying the point-like assumption of the Roche model. In practice with our modelling we saw,  by comparing Figs.~\ref{fig.1D_density_profile} and~\ref{fig.surface_deff_R_all_models}, that deformation discrepancies are directly related to the 1D mass profiles of each star.  Stars with the largest mass fraction in their upper layers are the most impacted by our modelling.  In particular, in our RGB model, the low density of the envelope is totally compensated by its significant volume, leading to an envelope mass of about 0.8 M$_{\mathrm{tot}}$ (see Fig. ~\ref{fig.1D_density_profile}).  MoBiDICT is necessary to model  deformations of close binaries composed of stars with high envelope mass fraction,  such as RGB or low-mass stars. 
\\~\\All the differences previously presented can have a significant impact on the theoretical evolution of binary stars,  as mass transfer will occur at higher orbital separation than predicted by the Roche model. 
Therefore, mass transfers and common envelope phases can occur in earlier system evolutionary stages than with the Roche model,  impacting all the future evolutionary path of such systems.  In that respect, the significant deformation discrepancy found for RGB stars is crucial as most of the mass transfers are expected to occur during late evolutionary stages due to the rapid radius inflation resulting from stellar evolution.  For example,  in the system presented in Fig.~\ref{fig.surface1M0}, a common envelope phase is expected to occur when the orbital period reaches 2h05min with our modelling,  while with the Roche model an orbital period of 2h is required.
 \\~\\In Sect~\ref{subsec_convergence_c} we mentioned and presented the different readjustments made at the end of each modelling iteration. One of these readjustments, presented in Appendix~\ref{annexe_period_readjustment},  is the computation of the system orbital period taking into account the non-spherical shapes of deformed binaries. In the case of low-mass binaries we saw an $\sim 1 \%$ increase of the orbital period for a given orbital separation. Compared to the usual uncertainties on the determination of orbital periods,  this effect dominates the uncertainties when linking the orbital period of a system to its orbital separation.

\subsection{Effective surface gravity}\label{subsec_surface_gravity_roche}
\begin{figure}[h]
\centering
\includegraphics[width=\hsize]{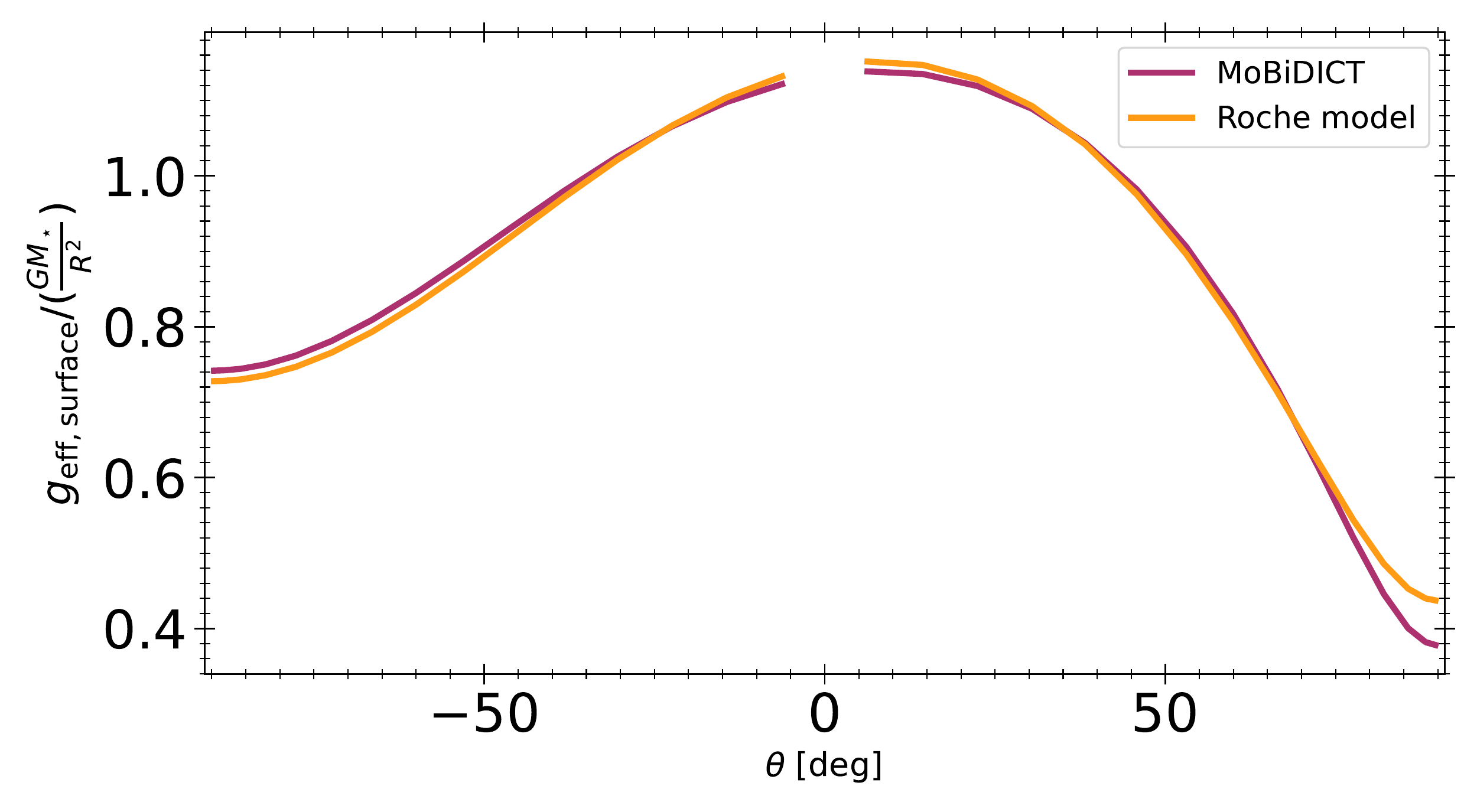}
\caption{ Surface effective gravity of the meridian passing by the most distorted point of the star,  its opposite,  and the pole.  The orange curve corresponds to the surface effective gravity given by the Roche model while the purple curve corresponds to the effective gravity obtained with our modelling.  The system modelled here is a twin system composed of two $0.2$ $M_\odot$ stars with an orbital separation of $2.8\ R_{\rm{1D}}$. The absence of points near $\theta=0$ is explained by the Gauss-Legendre quadrature method that is not creating any points in this region of our grids.  $\theta>0$ corresponds the direction where the stars are facing each other,  while the direction $\theta<0$ corresponds to the back of the system.}
\label{fig.surface_gravity_1M0}
\end{figure}

\begin{figure}[h]
\centering
\includegraphics[width=\hsize]{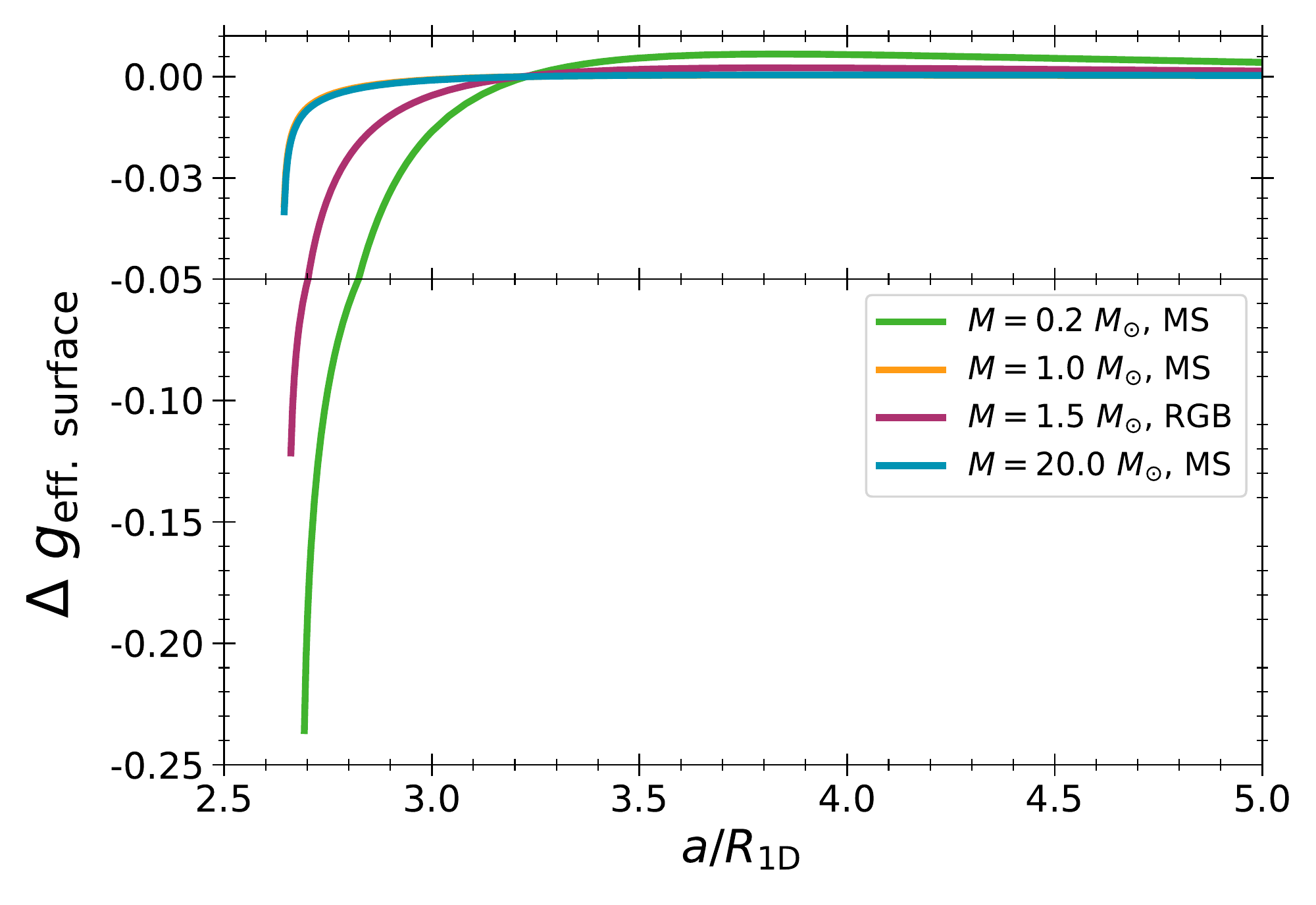}
\caption{ Difference of surface effective gravity of the most distorted point as a function of  orbital separation for the four types of twin binaries studied.}
\label{fig.surface_geff_all_models}
\end{figure}

On the observational side, more than having significantly larger deformations than Roche with MoBiDICT, the surface effective gravity is directly impacted as the topology of the total potential is modified.  The local brightness of a star as given,  for example,  in Eq.~\ref{eq_gravity_darkening} is proportional to the surface local effective gravity.  Therefore a modification of the total potential topology can directly be seen in the lightcurves of eclipsing binary systems and impact the derivation of the orbital parameters,  for example.  As we are going to focus on the surface effective gravity,  and both our refined Roche model and the classical Roche model give the same result,  only one of them is considered in the following of this section.
\\In our models the effective gravity has an integral expression given in Appendix \ref{annexe_effective_gravity},  and for the Roche model,  the effective gravity will be given by $\vert\textbf{g}_{\rm{eff}}\vert= \vert \boldsymbol{\nabla} \Psi_{\rm{tot,  Roche}}\vert$ where $\Psi_{\rm{tot,  Roche}}$ is given in Sect.~\ref{subsec_formulation_roche}.  Figure~\ref{fig.surface_gravity_1M0} compares, for the Roche model and our models,  the surface effective gravity along the meridian passing by the most distorted point and the poles of the stars for a twin system composed of two $0.2$ $M_\odot$ stars with an orbital separation of $2.8\ R_{\rm{1D}}$.
\\~\\Figure~\ref{fig.surface_gravity_1M0} shows that a difference of surface effective gravity exists between the Roche model and our modelling.  The most significant $g_{\rm{eff}}$ discrepancies are located a the closest region to the L1 point where we have a normalized $\rm{g}_{\rm{eff}}$ difference of $12\%$ with respect to the Roche model.  
\\The discrepancies found can be explained by several factors,  first, as our models are noticeably more distorted than the Roche model, the surface effective gravity is consequently significantly lower at the surface.  In a lesser extent,  not correcting the Roche model for the redistribution of mass in the star can impact the surface $g_{\rm{eff}}$.
\\To summarize the differences obtained,  we computed the effective gravity of the grid of models presented in Sect.~\ref{subsec_surface_deformation_roche} and quantified the differences of surface effective gravity between our models with the quantity $\Delta \rm{g}_{\rm{eff, surface}}$,  defined as 
\begin{equation}\resizebox{\hsize}{!}{%
$\Delta g_{\rm{eff}}=\frac{g_{\rm{eff,  MoBiDICT}}(\mu=0, \phi=0)-g_{\rm{eff,  Roche}}(\mu=0, \phi=0)}{GM_\star/R_{\rm{1D}}^2}.$}
\end{equation}
In Fig.~\ref{fig.surface_geff_all_models},  we summarized all the differences of surface effective gravity for the twin binary systems presented in Sect.~\ref{subsec_surface_deformation_roche} and for a wide diversity of orbital separations.
\\~\\Figure~\ref{fig.surface_geff_all_models} shows the same overall results than Fig.~\ref{fig.surface_deff_R_all_models}. The models with  most surface deformation discrepancies compared to the Roche model are the same models exhibiting the stronger surface effective gravity differences.   In particular,  Fig.~\ref{fig.surface_geff_all_models} shows that higher discrepancies of $\rm{g}_{\rm{eff}}$ are found for the RGB and low mass stars when $a/R_{\rm{1D}}\lesssim3.2$,  the other MS stars having limited corrections from our modelling. 
\\~\\For the interpretation of observations, the modifications of $\rm{g}_{\rm{eff}}$ found can, in theory,  strongly impact the lighcurves of close binary systems  as the local effective temperature of a star is directly proportional to the surface effective gravity.  With our modelling, we expect that a star would appear colder than with the Roche model.  In particular,  the equator is the region that would appear noticeably colder and thus redder.  To confirm these considerations,  simulations of binary lightcurves using models from MoBiDICT are the next step.

\subsection{Structural deformation}\label{subsec_structural_deformation_roche}
\begin{figure}[h]
\centering
\includegraphics[width=\hsize]{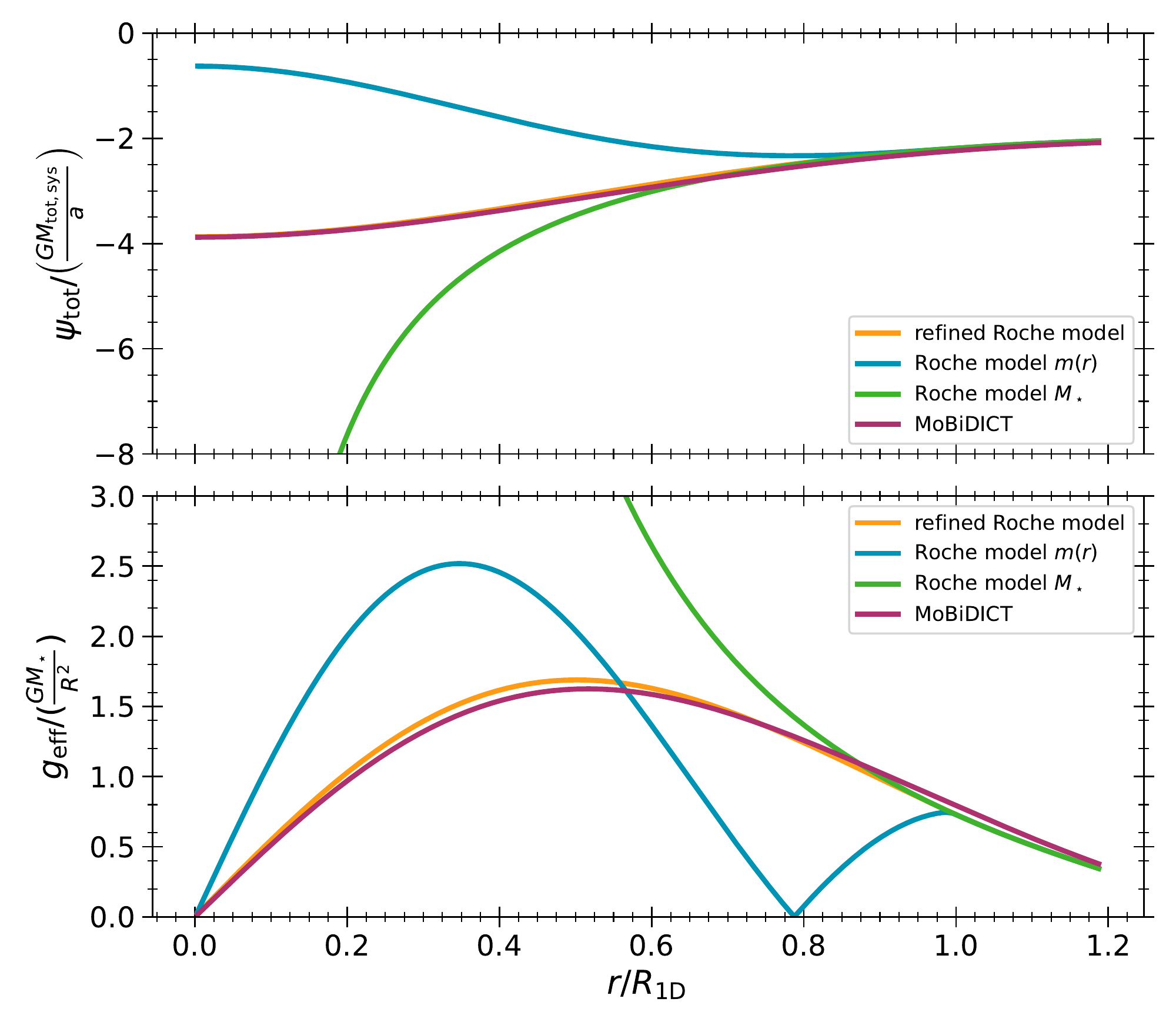}
\caption{Normalized total potential $\Psi_{\rm{tot}}$ and effective gravity $\rm{g}_{\rm{eff}}$  in the stellar interior along the axis joining the centre of the two stars. The system studied here is a twin binary system composed of $0.2 M_\odot$ stars with an orbital separation of $a=2.8R_{\rm{1D}}$. The blue and green curves are corresponding to the classical Roche model while the orange and violet curves are respectively our refined Roche model and our new types of models. }
\label{fig.pot_geff_roche}
\end{figure}

\begin{figure}[h]
\centering
\includegraphics[width=\hsize]{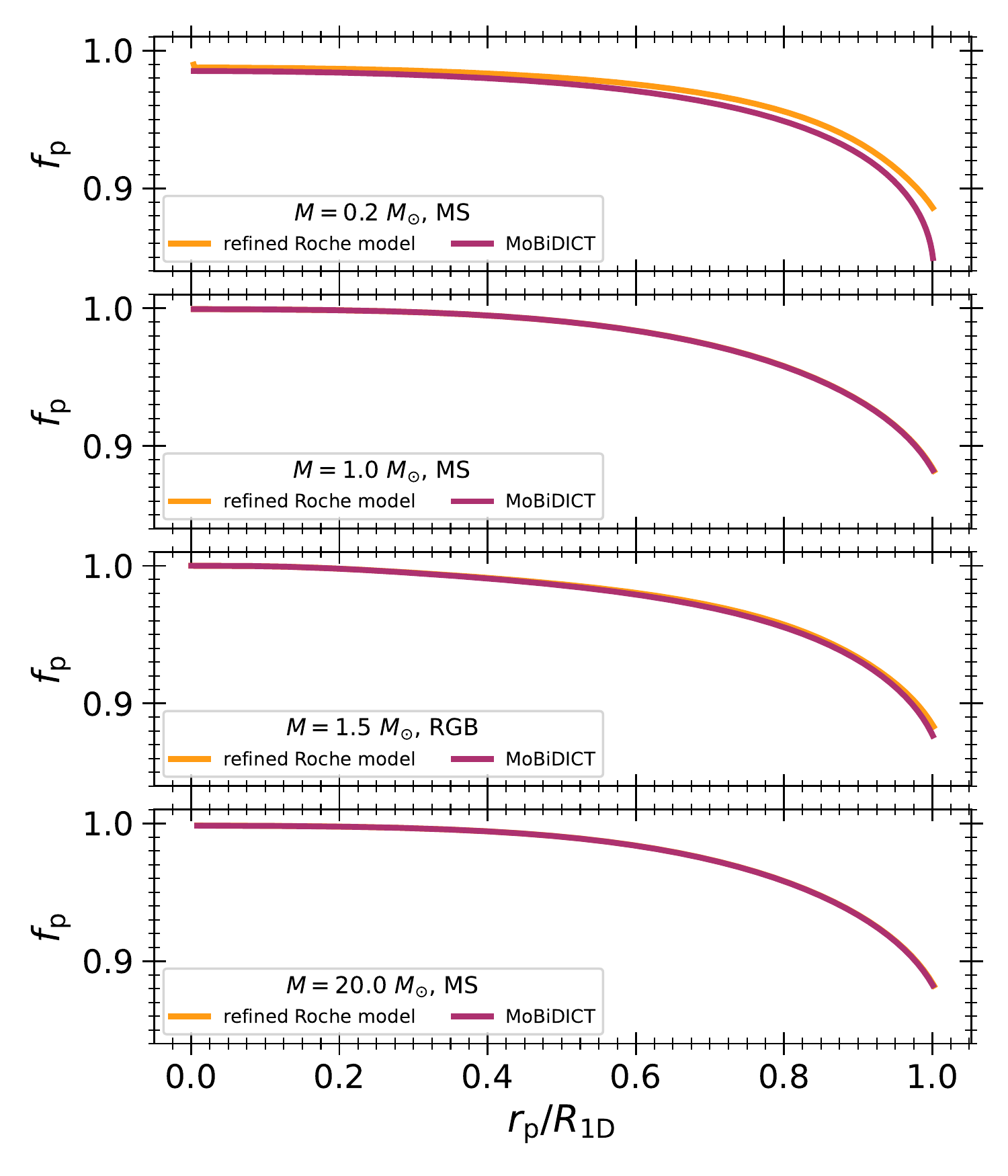}
\caption{Evolution of $f_{\rm{p}}$ as a function of the radius of stars in twin binary systems for the different models compared.  From top to bottom,  each panel respectively represents $ f_{\rm{p}}$ for the $0.2$,  $1.0$,  $1.5$ and $20.0$ $M_\odot$ stars in twin binary systems.  The orange curves are the results with our refined Roche model while the purple curves are the results from MoBiDICT.}
\label{fig.f_p_all_models}
\end{figure}

\begin{figure}[h]
\centering
\includegraphics[width=\hsize]{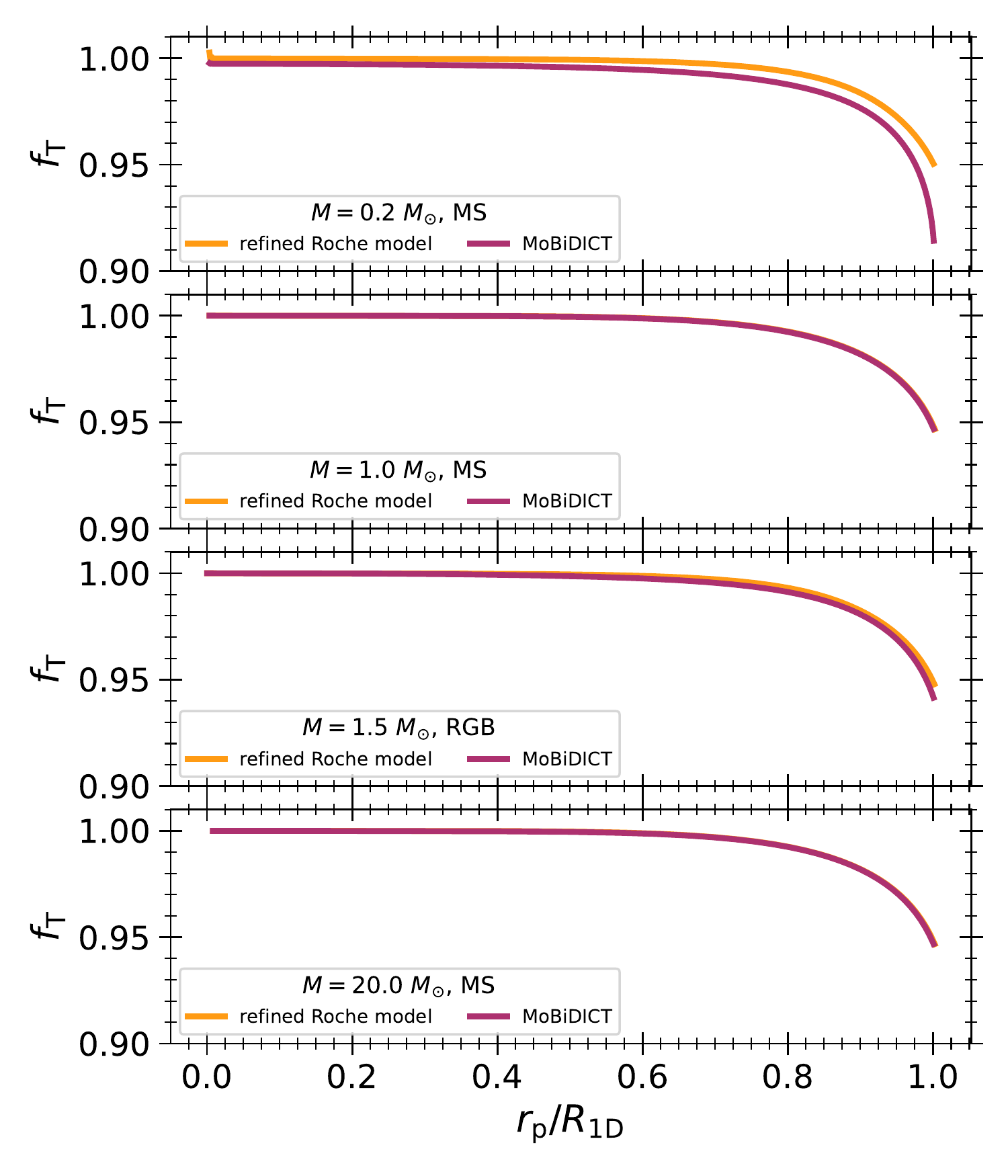}
\caption{ Evolution of $f_{\rm{T}}$ as a function of the radius of stars in twin binary systems for the different models compared.  From top to bottom,  each panel respectively represents $ f_{\rm{T}}$ for the $0.2$,  $1.0$,  $1.5$ and $20.0$ $M_\odot$ stars in twin binary systems.  The orange curves are the results with our refined Roche model while the purple curves are the results from MoBiDICT.}
\label{fig.f_t_all_models}
\end{figure}

The last point of comparison between the Roche model and our models is the structural deformation of the bodies composing the binary system.  For this study,  we compared the classical Roche model,  our refined Roche model and the model obtained through MoBiDICT. The structural discrepancy can be first seen in the total potential $\Psi_{\rm{tot}}$ and the effective gravity $\rm{g}_{\rm{eff}}$ as these quantities are defining the structural properties of our models. Figure~\ref{fig.pot_geff_roche} illustrates the evolution of $\Psi_{\rm{tot}}$ and $\rm{g}_{\rm{eff}}$ in a particular direction, the axis linking the centre of both stars and the system centre of mass for the $0.2 M_\odot$ stars with an orbital separation of $a=2.8R_{\rm{1D}}$. 
\\~\\Figure~\ref{fig.pot_geff_roche} illustrates the limitations of Roche models to study the stellar structure of deformed binaries. Classical ways of computing the Roche potential (Eqs.~\ref{eq_pot_roche_tot} and \ref{eq_pot_roche_simple}) are inaccurate to model stellar interior while their surface properties are reproduced. In addition, Fig~\ref{fig.pot_geff_roche} shows that our refined Roche model is particularity accurate to obtain both $\Psi_{\rm{tot}}$ and $\rm{g}_{\rm{eff}}$ even if discrepancies can be seen in the entire star.
\\In the work of \cite{Fabry2022}, the potential used to compute the $f_{\rm{p}}$ and $ f_{\rm{T}}$ factors,  Eqs.~(\ref{eq_fp}) and (\ref{eq_ft}),  is corresponding to the green curve in Fig.~\ref{fig.pot_geff_roche}, however, with assumptions on its derivative the corresponding effective gravity used is the expression given by our refined Roche model. Even if the effective gravity is correct to obtain $f_{\rm{p}}$ and $ f_{\rm{T}}$, the potential used to identify the equipotentials on which the average quantities are computed is not representative of the stellar deformed structure. Corrections can therefore be expected on the averaged internal quantities $f_{\rm{p}}$ and $ f_{\rm{T}}$ in \cite{Fabry2022},  for example. 
\\~\\To quantify the structural differences on the entire stellar structure,  we used the coefficients $f_{\rm{p}}$ and $ f_{\rm{T}}$ presented in Sect.~\ref{sec_post_treatment}.  
\\As we are considering the entire deformed structure as a function of the radius, we limited our study to twin binary systems composed of our four types of stars presented in Sect.~\ref{subsec_type_of_stars_studied} with the orbital separation of $2.7 \ R_{\rm{1D}}$. The resulting $f_{\rm{p}}$ and $ f_{\rm{T}}$ obtained with the different models are respectively illustrated in Fig.~\ref{fig.f_p_all_models} and Fig.~\ref{fig.f_t_all_models}. As classical expressions of the Roche model are leading to unphysical results on the majority of the stellar structure, $f_{\rm{p}}$ and $ f_{\rm{T}}$ are not studied for these models.
\\~\\Figures~\ref{fig.f_p_all_models} and \ref{fig.f_t_all_models}  show that models with the most deformation discrepancies  with respect to the refined Roche model are also the models with the higher $f_{\rm{p}}$ and $f_{\rm{T}}$ differences.  This structural difference can be mainly seen in upper layers of the $0.2$ and $1.5$ $M_\odot$ stars as the bodies have the largest deformation discrepancies compared to the Roche model.  Moreover,  in the case of the $0.2$ $M_\odot$ binary system,  a discrepancy in the entire structure can be seen. By predicting a lower $f_{\rm{p}}$ and $f_{\rm{T}}$,  our modelling is impacting the structure of binary stars. Combined to the discrepancies seen in deformations our modelling can significantly alter the evolutionary path of binary systems. Finally, as previously seen, for MS massive and solar like stars, our modelling is not significantly changing their structural depiction.

\section{Comparison with perturbative models}\label{sec_comparison_perturbative}

The principle of perturbative models is to treat the centrifugal and tidal forces as perturbation of the spherical symmetry only accounting for the leading terms.  The potentials and densities of the models are also projected on a basis of spherical harmonics only taking the leading order, namely the $\ell=0$ and $\ell=2$ terms.  The objective of this section is to determine what are the limits to only account for the leading orders in the spherical projections and to see whether the treatment of deformations by a perturbative approach is justified in the most distorted cases.
\\In Sect.~\ref{subsec_perturbative_formulaton},  we present the theoretical formulation of the perturbative models. Section~\ref{subsec_perturbative_potential} is dedicated to the exploration of the limitations of the selection of spherical harmonic projections of the gravitational potential,  while in Sect.~\ref{subsec_perturbative_apsidal} we study the limit of the perturbative approach in the treatment of the tidal and centrifugal forces. 

\subsection{Theoretical formulation of perturbative model}\label{subsec_perturbative_formulaton}
As explained previously, the perturbative models are a class of models treating the centrifugal and tidal forces as a first order perturbation to the spherical symmetry.  The gravitational potential is obtained, as in MoBiDICT, through Eq.~\ref{eq_poisson_solved_fidz} but only taking in account the first order terms: the $\ell=0$ and $\ell=2$. With this perturbative modelling,  the feedback process where the modification of the potential lead to a redistribution of the masses (therefore impacting the potential that requires to be updated in an iterative process) is not present.  Without going through the detailed demonstration of the equations used by the pertubative model,  we can mention that the assumptions made allowed to simplify the expression of the deformations that are related now to the structural coefficient $\eta_\ell$.  These coefficients are linked to the deformations of a star through their definition:
\begin{equation}\label{eq_definition_eta_j}
\eta_{\ell}^m(\overline{r})=\frac{\overline{r}}{r_{\ell}^m(\overline{r})}\frac{\mathrm{d} r_{\ell}^m(\overline{r})}{\mathrm{d} \overline{r}},
\end{equation}
where $\overline{r}$ is the average radius of an isobar that is defined as
\begin{equation}
\overline{r}=\frac{1}{\int_{\Psi_{\rm{tot}}} \mathrm{d}\sigma} \int_{\Psi_{\rm{tot}}} r(\mu,\phi) \mathrm{d}\sigma,
\end{equation}
and that is different from $r_{\rm{p}}$ corresponding to the radius of the sphere with the identical volume than the one under an isobar. $r_{\ell}^m$ are the radii of the equipotentials $ r(\mu, \phi)$ projected on the spherical harmonics basis as done in Eq.~\ref{eq_rho_lm},  that is expressed in this case as
\begin{equation}\label{eq_projection_deformation}
r_{\ell}^m(\overline{r})=\frac{1}{\overline{r}}\int_0^1\int_0^\pi r(\overline{r},\mu, \phi) P_{\ell}^m (\mu) \cos(m\phi) \mathrm{d}\phi \mathrm{d}\mu.
\end{equation}
The overall deformation of a given point in a model can thus be obtained by projecting back these $r_{\ell}^m(\overline{r})$ on the spherical coordinates basis as follows
\begin{equation}
r(\overline{r},\mu, \phi)=\overline{r}\sum^{(\ell-p)/2}_{k=0} r_{\ell}^m(\overline{r}) P_{\ell}^m (\mu) \cos(m\phi),
\end{equation}
with  $p=0$ if $\ell$ is even, $p=1$ otherwise,  $m=2k+p$. 
\\With the coefficient $\eta_\ell$ and using a simple integration,  $r_{\ell}^m(\overline{r})$ can be obtained,  and thus,  the deformation of a model can be computed.  Physically,  these coefficient $\eta_\ell$ are corresponding to the structural answer of a body,  if a perturbative potential is applied.  Using simplifications done by the perturbative approach,  $\eta_\ell$ are becoming independent from the deformation undergone by a star and only related to it's unperturbed structure.  $\eta_\ell$  can be obtained, in the perturbative approach,  by solving the well known Clairaut-Radau second order differential equation expressed as
\begin{equation}\label{eq_clairaut-radau}
\overline{r}\frac{\mathrm{d} \eta_{\ell}}{\mathrm{d}\overline{r}}+6\frac{\rho(\overline{r})}{\overline{\rho}(\overline{r})}(\eta_{\ell}+1)+\eta_{\ell}(\eta_{\ell}-1)=\ell(\ell+1),
\end{equation}
using the boundary conditions
\begin{equation}
\eta_\ell(0)=\ell-2.
\end{equation}
To compute the perturbative models,  we developed a small feature in MoBiDICT to solve the Clairaut-Radau equation (Eq.~\ref{eq_clairaut-radau}) with an order two Rung-Kutta method,  using the 1D input stellar models given in this tool.  
\\For the comparison with model obtained with MoBiDICT,  we decided to directly compare the $\eta_\ell$ of the two models. To obtain these   coefficients with our models,  we compute the deformations as described in Sect.~\ref{sec_stellar_models}. Then we obtained $\overline{r}$ and projected the deformations on a basis of spherical harmonics (Eq.\ref{eq_projection_deformation}) to have $r_{\ell}^m(\overline{r})$ that were used in Eq.\ref{eq_definition_eta_j} to compute the $\eta_{\ell}^m(\overline{r})$ coefficients directly given by MoBiDICT. We can mention that passing by these equations,  none of the assumptions of the perturbative approach was made.

\subsection{Gravitational potential}\label{subsec_perturbative_potential}
\begin{figure}[h]
\centering
\includegraphics[width=\hsize]{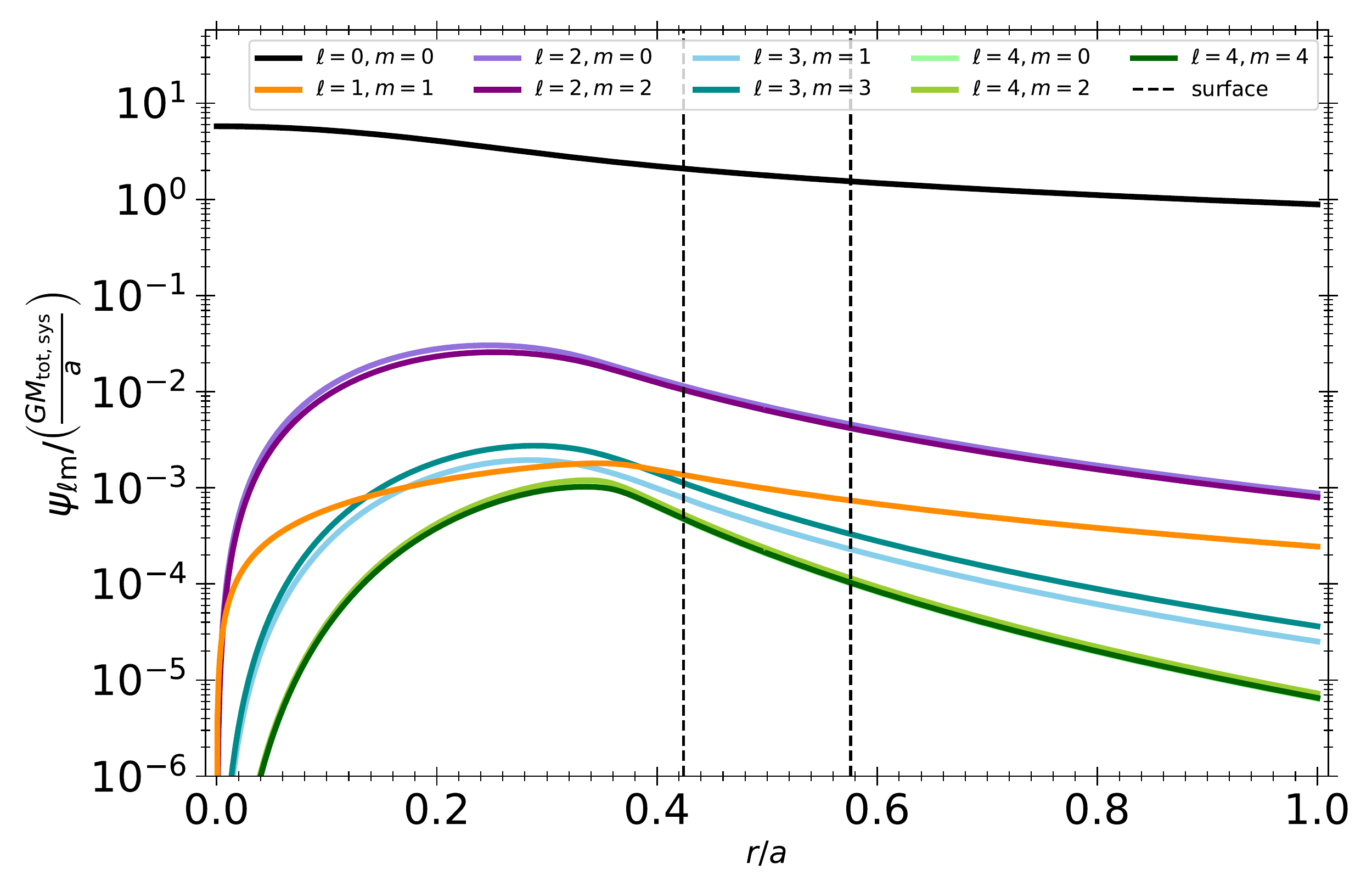}
\caption{Normalized gravitational potential projected on the spherical harmonics of a star of $0.2\ M_\odot$ in a twin binary system with an orbital separation of $a= 2.8R_{\rm{1D}}$. The dotted lines are corresponding to the surface of each star of the system in the direction including the most distorted point of each star and the Lagragian point L1 ($\mu=0, \phi=0$).}
\label{fig.spectral_potential}
\end{figure}

One main approximation of perturbative models is to only account for the leading order terms ($\ell=0$ and $\ell=2$) on the projections of the potential and densities. Verifying these assumptions with MoBiDICT is straightforward as our modelling is directly giving the spherical expansion of the gravitational potential of each deformed star as an output of the code. Figure~\ref{fig.spectral_potential} illustrates the gravitational potential projected on the spherical harmonics basis of the twin binary system composed of the $0.2\ M_\odot$ stars at an orbital distance of $a= 2.8R_{\rm{1D}}$, this is the exact system  configuration illustrated in Fig.\ref{fig.surface1M0}.
\\~\\Figure~\ref{fig.spectral_potential} illustrates the amplitude hierarchy of gravitational potential spherical terms. As assumed by the perturbative model,  the leading order terms are the $\ell=0$ and $\ell=2$ terms.  The dipolar term ($\ell=1$) is the next most important term, in particular outside of the star as each spectral term is proportional to $r^{-\ell}$.  In these regions the dipolar term is included in the computation of the tidal potential applied to the opposite star. Neglected by the perturbative model, this $\ell=1$ term is less than one order of magnitude smaller than the $\ell=2$ term, meaning that the tidal force is significantly greater with our models and the assumption to neglect the $\ell=1$  is not justified for the perturbative models in the most distorted cases. The next order terms ($\ell=3,4$) have similar amplitudes than the $\ell=1$ term inside the star,  while they are one order of magnitude smaller than this same dipolar term outside the star.

\subsection{Validity of the Clairaut-Radau equation}\label{subsec_perturbative_apsidal}
\begin{figure}[h]
\centering
\includegraphics[width=\hsize]{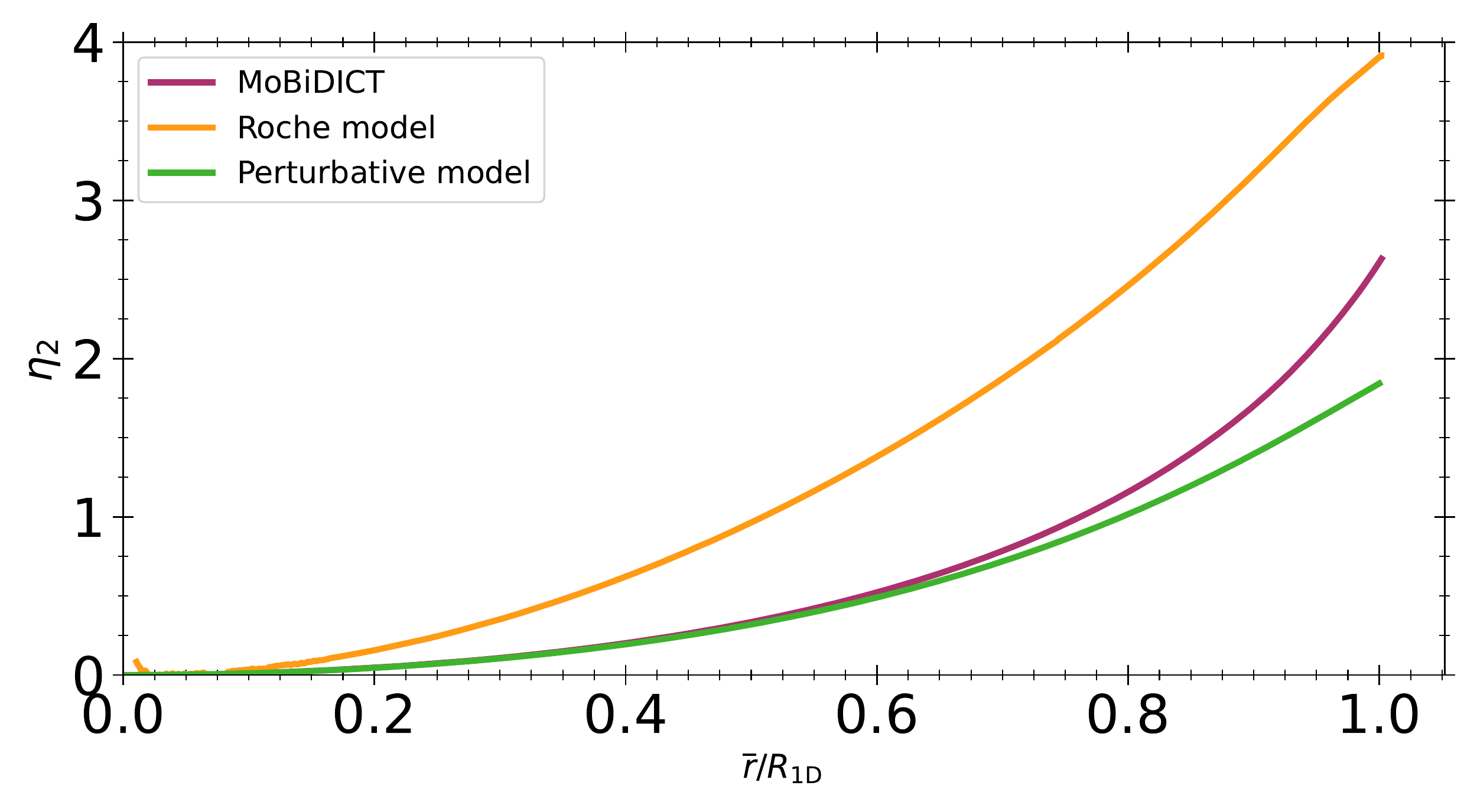}
\caption{Evolution of $\eta_2$ as a function of the average radius for the different models studied in this work.  The system studied here is a twin binary system composed of two  $0.2\ M_\odot$ stars with an orbital distance of $2.8R_{\rm{1D}}$. }
\label{fig.eta2_evolution}
\end{figure}

\begin{figure}[h]
\centering
\includegraphics[width=\hsize]{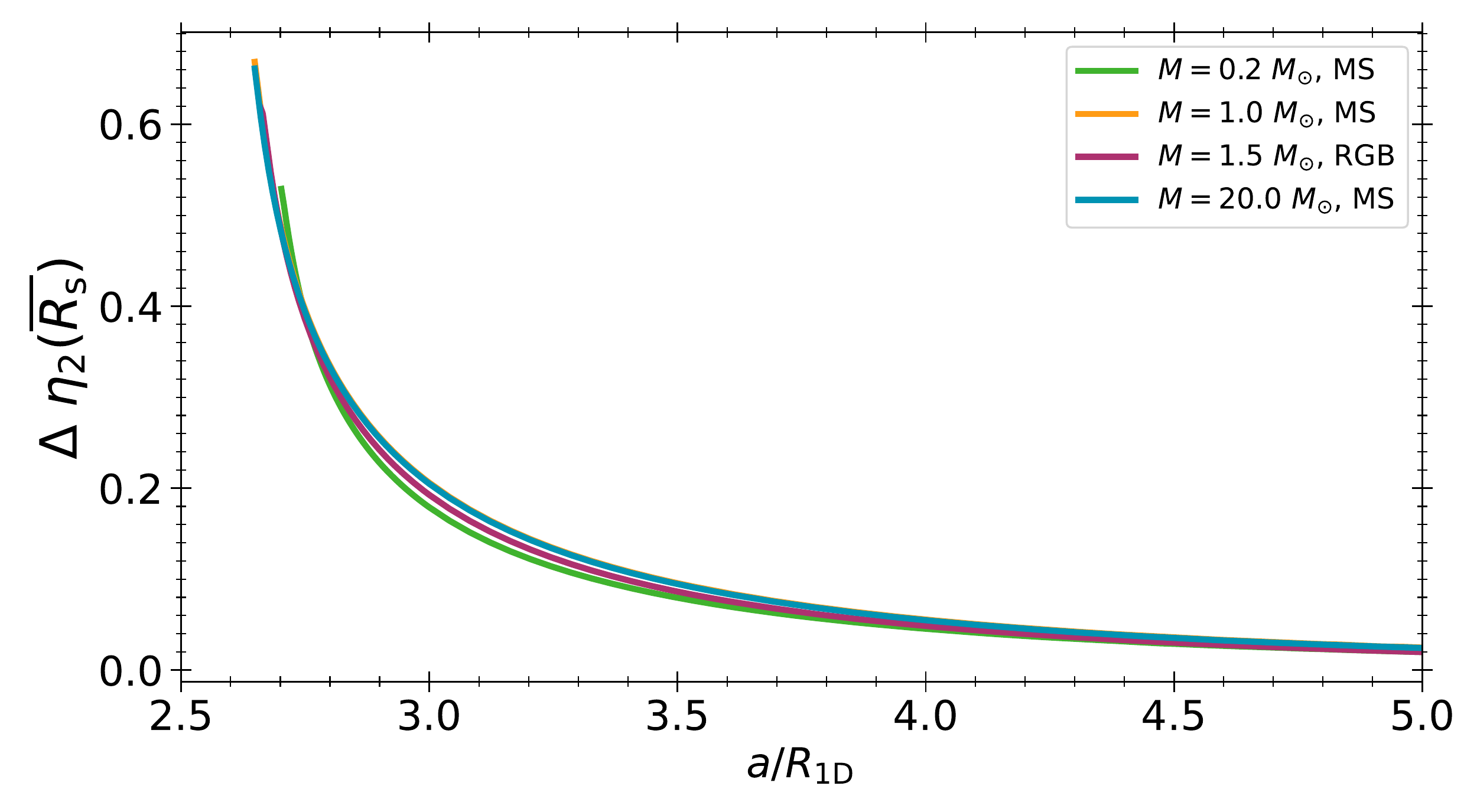}
\caption{ Evolution of the surface $\Delta \eta_2$ for different types of stars composing the binaries and with different orbital separation.}
\label{fig.summar_eta2}
\end{figure}

In this section,  we verify the limits of the perturbative approach for close binaries, in particular when the stars are undergoing high deformations.  To compare the different models, as explained in Sect.~\ref{subsec_perturbative_formulaton},  we directly look at the coefficient $\eta_\ell$ obtained through Eq.~\ref{eq_clairaut-radau} for perturbative models,  while with our models theses quantities are computed with a projection on the spherical harmonics of the deformations.  More details are given in Sect.~\ref{subsec_perturbative_formulaton}.  We also compared these models to our refined Roche model through the same coefficients $\eta_\ell$ obtained by the same way that with our models. We focus our work on the $\ell=2$ term as this coefficient is assumed, by the perturbative model,  to be the leading contribution to the aspidal motion of binary system that is used for example in \cite{Rosu2020, Rosu2022a, Rosu2022b} to constrain the stellar structure.
\\The evolution of $\eta_2$ as a function of the average radius is illustrated in Fig.~\ref{fig.eta2_evolution} for the extensively studied system of this article,  the twin binary system composed of $0.2\ M_\odot$ stars at an orbital distance of $2.8R_{\rm{1D}}$.
\\~\\Figure~\ref{fig.eta2_evolution}  illustrates different significant results from our modelling. First comparing the Roche model to other models,  we see that both perturbative  and MoBiDICT models are predicting an important difference of $\eta_2$ in all the stars.  These results can be interpreted as difficulties of the Roche model to precisely depict the stellar deformations, these limitations originating from the assumptions on the stellar gravitational potential. 
\\The second major result illustrated in Fig.~\ref{fig.eta2_evolution} is the comparison between our new models and the perturbative models.  First, in the less distorted regions of the star (when $\overline{r}<0.5 R_{\rm{1D}}$) the same results are obtained with models,  indicating that our modelling is giving a good depiction of stellar deformation in these regions,  as we expect to be in the limit of validity of perturbative modelling. Then, after $\overline{r}>0.5 R_{\rm{1D}}$,  non negligible differences between the two models are appearing,  with discrepancies increasing with $\overline{r}$,  as the upper layers of the star are more deformed. These differences can be explained by the limitations of the pertubative approach in these regimes of deformation,  while our model is designed to study cases with high stellar deformations. 
\\~\\To study the apsidal motion of binary system one would require the apsidal motion constant $k_2$ that is usually obtained through $\eta_2$ evaluated at the surface of the primary star.  To study how our new modelling procedure is impacting this apsidal motion constant depending on the twin binary system studied and orbital separation we introduce a new quantity $\Delta \eta_2$ defined as the difference of $\eta_2$ at the stellar surface between our model and the perturbative model,  $\Delta \eta_2$ is expressed as
\begin{equation}
\Delta \eta_2=\eta_{2, \rm{MoBiDICT}}(\overline{R}_{\rm{s}})-\eta_{2, \rm{pert.}}(\overline{R}_{\rm{s}}).
\end{equation}
The comparison of $\Delta \eta_2$ from the grid of twin binary models, presented in Sect.~\ref{subsec_type_of_stars_studied}, as a function of their orbital separation,  is illustrated in Fig.~\ref{fig.summar_eta2}.
\\~\\Figure~\ref{fig.summar_eta2} illustrates that the differences of $\eta_2$ between the perturbative and MoBiDICT models are almost independent from the type of star composing the binary system and only dependent on the orbital separation of the components. Significant differences of surface $\eta_2$ can only be seen in the most distorted cases, when $a<5R_{\rm{1D}}$,  otherwise the perturbative treatment of the tidal and centrifugal forces is justified to compute $\eta_2$.  The differences found are expected to directly impact the apsidal motion computation, predicting correction to apsidal motion constant $k_2$ and adding $k_1$ constant to account for neglected dipolar terms. The impact of our improved modeling on apsidal motion will be deeper investigated in another forthcoming article. 

\section{Conclusion}\label{sec_conclusion}

In this work we developed a new tool called MoBiDICT to compute 3D static models of close, synchronized binaries in hydrostatic equilibrium in a non-perturbative way. Sections \ref{sec_stellar_models} and \ref{sec_post_treatment} were respectively dedicated to give a technical description of this code and present the post treatment done in our models to couple, in the future, MoBiDICT to classical 1D evolution codes. Then, we compared our new type of modelling to classical models, namely the Roche and perturbative models.
\\~\\First,  for the Roche model,  our study showed in Sect.\ref{sec_comparison_Roche} that the differences of deformation are mainly arising in very close binaries (when $a<5R_{\rm{1D}}$).  The discrepancies are significant for low mass and RGB stars in twin binary systems where,  the deformation differences can reach at most $80\%$ for our $0.2M_\odot$ stars.  Massive and solar-like MS stars are less distorted,  with a difference of $7\%$ at most.  The envelope mass is the key parameter controlling these deformation discrepancies.  Stellar models with the largest envelope masses compared to the total mass are the most modified by MoBiDICT. All the quantities studied were impacted by this difference, in particular significant changes of the surface effective gravity were seen directly affecting gravity darkening and thus the interpretation of the observations of such stars. In the lower deformation regime,  the Roche model is a good approximation of the stellar surface properties while we would not favour Roche to study the structural properties of stars.
\\~\\The discrepancies with respect to the perturbative approach were finally discussed in Sect.~\ref{sec_comparison_perturbative}. Our work showed that the assumption of the perturbative model to only account for the leading terms of the projected gravitational potential ($\ell=0$ and $\ell=2$) is not justified in the high distortion cases (when $a<5R_{\rm{1D}}$).  In particular, the dipolar term ($\ell=1$) was representing $31\%$ of the quadrupolar term ($\ell=2$) outside of star studied and $14\%$ at its surface.  In addition, we showed that in such regimes of distortion the perturbative approach reaches its limit and significant deformation discrepancies can be seen, in particular in the upper layers of the stars (when $r>0.5R_{\rm{1D}}$).  These differences of deformation are seen through the quantity $\eta_2$ that is evaluated at the surface to estimate the apsidal motion of binary systems.  In such distortion regime the theoretical apsidal motion from the quadroplar term is generally considerably underestimated,  and an additional term from the dipolar component is likely to arise.  In the least distorted regime,  ($a>5R_{\rm{1D}}$) the perturbative approach remains valid, and accurate $\eta_2$ were obtained.
\\~\\A few conclusions can be drawn from our work, first, in least distorted regimes the classical modelling methods for binaries are valid,  at the surface for the Roche model and in the entire structure for perturbative models. In the most distorted regimes, however, both perturbative and Roche models are failing to describe the structural and surface properties of stars with high envelope masses. In particular,  significant deformation discrepancies are observed for low mass stars, in agreement with the results of \cite{Landin2009}. Our work highlights the necessity of non-perturbative modeling to study close binary interactions and evolution, as mass transfer is expected to occur earlier in the lifetime of a system under our modeling approach.  In that regard,  deformation discrepancies found for RGB stars are significantly facilitating mass transfer occurrence during a stellar evolutionary phase already believed to be a source of mass transfer.
\\~\\Our method, while effective in modeling highly distorted cases, has demonstrated limitations when deformations are extremely weak. Specifically, we have observed numerical noise arising from higher spherical orders of the potential and densities,  which impact the derived $\eta_{\ell}$.   To address this issue, we plan to implement a coupling of MoBiDICT, in the least distorted regions, to a refined perturbative approach accounting for all the spherical orders and for terms of the tidal and centrifugal potential that are usually neglected. The advantage of this new formalism of the perturbative model is the possibility to study solid body non-synchronized systems that we are also able to easily generalize in MoBiDICT.  All these modifications will be the subject of a forthcoming article.

\begin{acknowledgements} The authors are thanking the anonymous referee for his/her constructive comments. 
L.F was supported by the Fonds de la Recherche Scientifique FNRS as a Reaserch Fellow. 
The authors would like to thank Martin Farnir from "Psychedelic Kiss Designs" for his help in finding our code name and designing  MoBiDICT's logo. 
\begin{figure}[h]
\centering
\includegraphics[width=4cm]{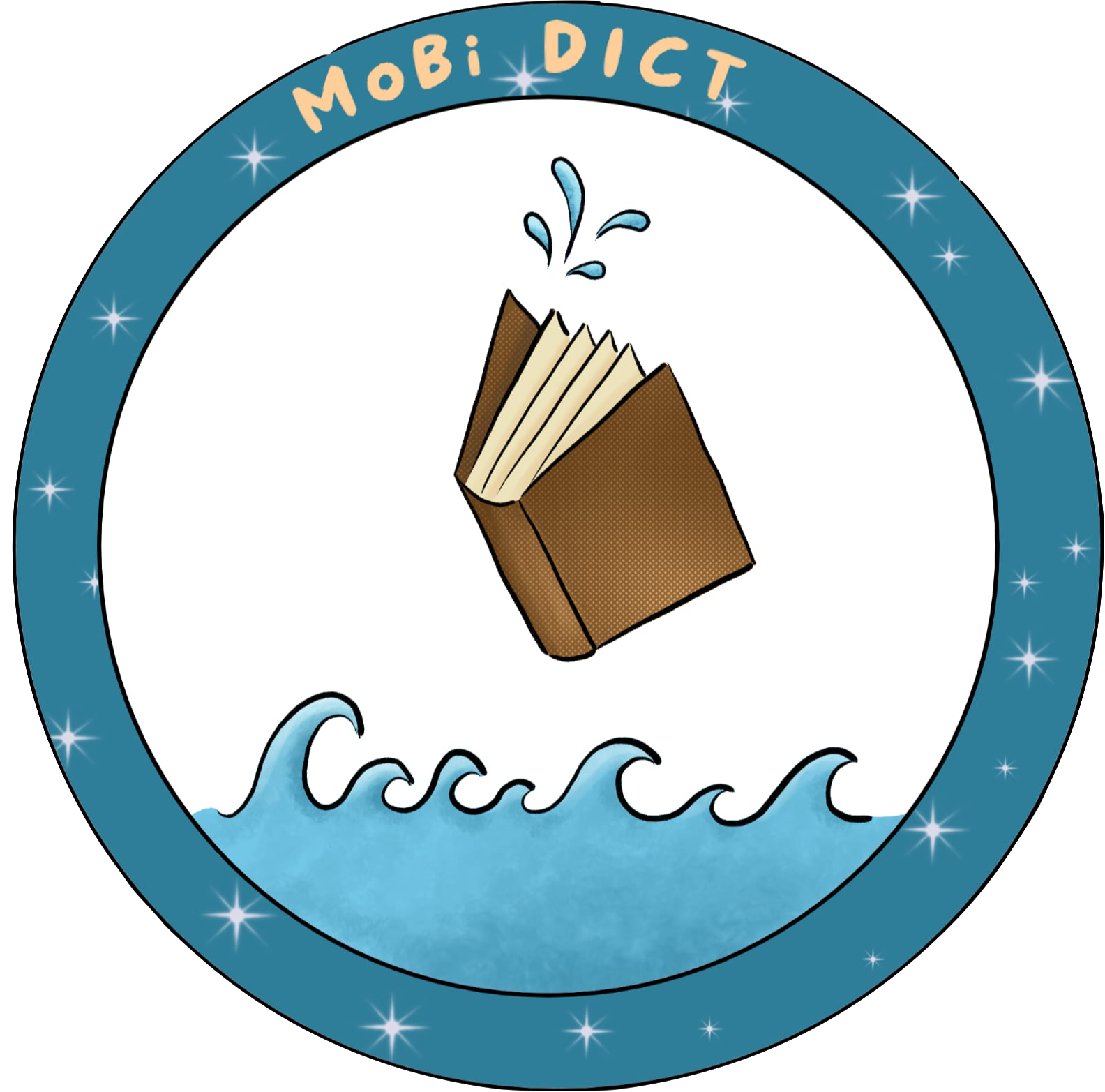}
\end{figure}
\end{acknowledgements}

\bibliographystyle{aa}
\bibliography{Article_1_bib}

\clearpage
\appendix

\section{Verification of the 3$^{\rm{rd}}$ Kepler's law}\label{annexe_period_readjustment}
\begin{figure}[h]
\centering
\includegraphics[width=\hsize]{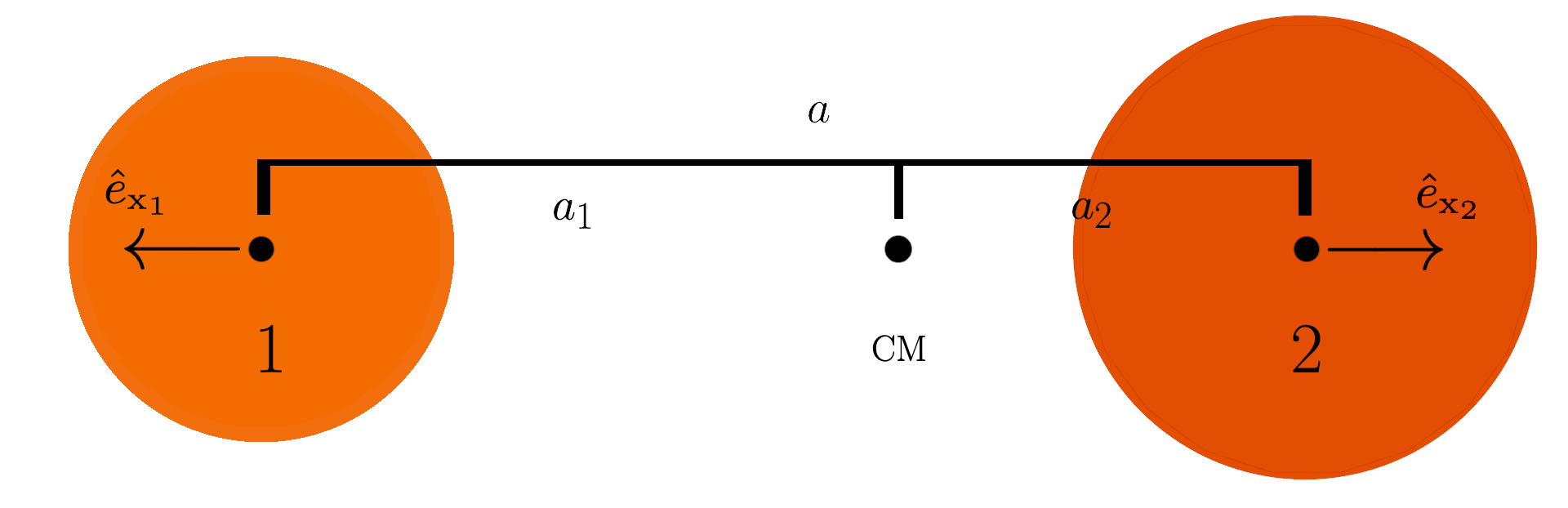}
\caption{Schema of an arbitrary chosen binary system composed of two stars with different masses.  $a_1$ and $a_2$  are the orbital separation of respectively the primary and secondary star from the centre of mass.  $\hat{e}_{\rm{x}_1}$ and $\hat{e}_{\rm{x}_2}$ are the unit vectors of the $\rm{x}_1$ and $\rm{x}_2$ direction considered in this problem. }
\label{fig.system_representation}
\end{figure}

In this section,  we show the system of equations solved at the end of each modelling iteration to readjust the centre of mass and the orbital period of our models.
\\Let us consider an arbitrary synchronized binary system with an orbital separation $a$,  an orbital rotation rate $\Omega_\star$ and a centre of mass noted $\rm{CM}$. This system can be illustrated as in Fig.~\ref{fig.system_representation}.
\\The orbital separation of the system $a$ is a constant of the model,  thus the centre of each star is at the dynamical equilibrium meaning that the sum of the gravitational,  tidal and centrifugal forces is null in their centre.  Mathematically this condition is expressed as 
\begin{equation}\label{eq.sys_first_step}
    \begin{cases}
      \lim_{\rm{x}_1 \rightarrow 0} \dfrac{\partial \Psi_1}{\partial \rm{x}_1}+\dfrac{\partial \Psi_2}{\partial \rm{x}_1}=\Omega_\star^2 x_{\rm{CM}} a   \\\\
      \lim_{\rm{x}_2 \rightarrow 0} \dfrac{\partial \Psi_1}{\partial \rm{x}_2}+\dfrac{\partial \Psi_2}{\partial \rm{x}_2}=\Omega_\star^2 (1-x_{\rm{CM}}) a
    \end{cases},
\end{equation}
where $x_{\rm{CM}}$ is defined as $a_1/a$.
\\The system given in Eq.~\ref{eq.sys_first_step} can be summed and rewritten as 
\begin{equation} \resizebox{\hsize}{!}{$
    \begin{cases}
      \lim_{\rm{x}_1 \rightarrow 0}  \dfrac{\partial \Psi_1}{\partial \rm{x}_1}+\dfrac{\partial \Psi_2}{\partial \rm{x}_1}=\Omega_\star^2 x_{\rm{CM}} a \\\\
            \lim_{\rm{x}_1 \rightarrow 0} \left(\dfrac{\partial \Psi_1}{\partial \rm{x}_1}+\dfrac{\partial \Psi_2}{\partial \rm{x}_1}\right)+\lim_{\rm{x}_2 \rightarrow 0} \left(\dfrac{\partial \Psi_1}{\partial \rm{x}_2}+\dfrac{\partial \Psi_2}{\partial \rm{x}_2}\right)=\Omega_\star^2 a   
    \end{cases}. $}
\end{equation}
The second equation of this system can be solved with MoBiDICT to obtain the $\Omega_\star^2$ necessary to maintain the orbital separation of the binary system.  Having $\Omega_\star^2$,  the first equation can be solved to obtain the new position of the centre of mass of the system accounting for the redistribution of mass in the stars.  

\section{Expression of the effective gravity}\label{annexe_effective_gravity}
In this section we are developing on the expression of the effective gravity used in MoBIDICT. 
\\The effective gravity is a vector defined by
\begin{dmath}
\textbf{g}_{\mathrm{eff}}=-\boldsymbol{\nabla}\Psi_{\rm{tot}}=-\boldsymbol{\nabla}\left(\Psi_1 -\Psi_2 - \Psi_{\rm{centri}}\right)=-(g_{\mathrm{eff}, r};g_{\mathrm{eff}, \mu}; g_{\mathrm{eff}, \phi})
\end{dmath} 
with, for a given primary star noted star 1, each component expressed as
\begin{equation}
g_{\mathrm{eff} r, 1}=\dfrac{\partial \Psi_{\rm{tot,1}}}{\partial r_1}= \dfrac{\partial \Psi_1}{\partial r_1}+\dfrac{\partial \Psi_2}{\partial r_1}+\dfrac{\partial \Psi_{\rm{centri}}}{\partial r_1}, 
\end{equation}
\begin{equation}
g_{\mathrm{eff}, \mu, 1}=\frac{1}{r_1}\dfrac{\partial \Psi_{\rm{tot,1}}}{\partial \theta_1}=\dfrac{1}{r_1}\left(\dfrac{\partial \Psi_1}{\partial \theta_1}+\dfrac{\partial \Psi_2}{\partial \theta_1}+\dfrac{\partial \Psi_{\rm{centri}}}{\partial \theta_1}\right), 
\end{equation}
\begin{dmath}
g_{\mathrm{eff}, \phi, 1}=\frac{1}{r_1\sin(\phi_1)}\dfrac{\partial \Psi_{\rm{tot,1}}}{\partial \phi_1}= \frac{1}{r_1\sin(\theta_1)}\left(\frac{\partial \Psi_1}{\partial \phi_1}+\frac{\partial \Psi_2}{\partial \phi_1}+\frac{\partial \Psi_{\mathrm{centri}}}{\partial \phi_1}\right).
\end{dmath}
In this case,  all the necessary quantities are partial derivatives of solutions to differential equations, which have integral representations. Consequently,  numerical methods are not required to perform the derivatives of the potentials. The contribution to effective gravity from gravitational potential of the primary star is given by 
\begin{align}
\frac{\partial \Psi_1}{\partial r_1} &=\sum_{\ell, m} P_{\ell}^m (\mu_1) \cos(m\phi_1)\frac{\partial \Psi_{1, \ell}^m}{\partial r_1} \\ &= 
\sum_{\ell, m} P_{\ell}^m (\mu_1) \cos(m\phi_1) \left( \Psi_{1, \ell}^m \frac{\ell}{r_1} \right.\\ &+ \left.  r_1^{-(\ell+2)}\int^{r_{1}}_{0} r_{1}^{\prime\ell+2} \rho_{1, \ell}^m(r_1^{\prime}) \mathrm{d}r^{\prime} \right)\nonumber,
\end{align}

\begin{dmath}
\frac{\partial \Psi_1}{\partial \theta_1}=-\sum_{\ell, m} \Psi_{1, \ell}^m(r_1) \cos(m\phi_1)\frac{\partial P_{\ell}^m (\mu_1)}{\partial \mu_1}  \sin(\theta_1),
\end{dmath}

\begin{dmath}
\frac{\partial \Psi_1}{\partial \phi_1}=-\sum_{\ell, m} \Psi_{1, \ell}^m(r_1) m\sin(m\phi_1) P_{\ell}^m (\mu_1).
\end{dmath}

The contribution from centrifugal force is expressed as
\begin{align}
\boldsymbol{\nabla}\Psi_{\rm{centri}}
&= \Omega_{\star}^2 \left( x_{\rm{CM}}-r\sin(\theta)\cos(\phi)\right)
\begin{bmatrix}
\sin(\theta)\cos(\phi)  \\\\
r\cos(\theta)\cos(\phi)  \\\\
-r\sin(\theta)\sin(\phi) \\
\end{bmatrix}\\ &-\Omega_{\star}^2 \left(r\sin(\theta)\sin(\phi)\right)
\begin{bmatrix}
\sin(\theta)\sin(\phi)  \\\\
r\cos(\theta)\sin(\phi)  \\\\
r\sin(\theta)\cos(\phi) \\
\end{bmatrix}\nonumber
,
\end{align}

Finally, the gravitational contribution originating from the companion is written as

\begin{multline}
\begin{bmatrix}
\dfrac{\partial \Psi_2}{\partial r_1} \\\\
\dfrac{\partial \Psi_2}{\partial \theta_1} \\\\
\dfrac{\partial \Psi_2}{\partial \phi_1} \\\\
\end{bmatrix} 
= -\textbf{J}^{-1}\\ \sum_{\ell, m}  \Psi_{2, \ell}^m(R_{0,2})\left(\frac{R_{0,2}}{r_2}\right)^{\ell+1}
\begin{bmatrix}
\dfrac{\ell +1}{r_2}P_{\ell}^m (\mu_2) \cos(m\phi_2) \\\\
\dfrac{\partial P_{\ell}^m (\mu_2)}{\partial \mu_2}\sin(\theta_2) \cos(m\phi_2) \\\\
P_{\ell}^m (\mu_2)m\sin(m\phi_2) 
\end{bmatrix},
\end{multline}
where $\textbf{J}^{-1}$ is the inverse of the jacobian matrix taken between the system of coordinates of our primary and secondary star,  and expressed as
\begin{equation}
\textbf{J}^{-1}=
\begin{bmatrix}
\dfrac{\partial r_2}{\partial r_1} & \dfrac{\partial \theta_2}{\partial r_1} & \dfrac{\partial \phi_2}{\partial r_1}\\ \\
\dfrac{\partial r_2}{\partial \theta_1} & \dfrac{\partial \theta_2}{\partial \theta_1} & \dfrac{\partial \phi_2}{\partial \theta_1}\\\\
\dfrac{\partial r_2}{\partial \phi_1} & \dfrac{\partial \theta_2}{\partial \phi_1} & \dfrac{\partial \phi_2}{\partial \phi_1}\\
\end{bmatrix}.
\end{equation}
The norm of the effective gravity is simply given by 
\begin{equation}
\vert\textbf{g}_{\rm{eff}}\vert= (g_{\rm{eff}, r, 1}^2+g_{\mathrm{eff}, \mu, 1}^2+g_{\mathrm{eff}, \phi, 1}^2)^{1/2} .
\end{equation}

\end{document}